\documentclass[11pt]{article}
\usepackage{epsfig}
\usepackage{graphicx}
\usepackage{amsmath}
\usepackage{amssymb}
\usepackage{authblk}

\newcommand{\Bf}{\boldsymbol{B}}
\newcommand{\Ef}{\boldsymbol{E}}
\newcommand{\rperp}{\rho}

\begin{document}

\title{Solar modulations by the regular \\
heliospheric electromagnetic field}

\author{Paolo Lipari}

\affil{\footnotesize 
INFN, sezione di Roma, Piazzale Aldo Moro 2, 00185 Roma, Italy.}

\date{August 2, 2014}

\maketitle
\begin{abstract}
The standard way to model the cosmic ray solar modulations
is via the Parker equation, that is as the effect 
of diffusion in the turbulent magnetic of 
an expanding solar wind.
Calculations performed with this method
that do not include a description of the regular magnetic
field in the heliosphere predict, in disagreement with the observations,
equal modulations for particles and antiparticles.
The effects of the regular heliospheric field, 
that break the the particle/anti--particle symmetry, have been included 
in the Parker equation adding convection terms 
associated to the magnetic drift velocity of
charged particles moving in non--homogeneous magnetic field.
In this work we take a completely different approach
and study the propagation of charged particles in the heliosphere 
assuming only the presence of the regular magnetic field,
and completely neglecting the random component.
Assuming that the field is purely magnetic in the wind frame,
one can deduce the existence of a large scale electric field, 
that results in important energy losses for charged particles that
traverse the heliosphere and reach the Earth.
The energy loss $\Delta E$ due to the large scale regular electric field
depends on the particle electric charge,
is proportional to the absolute value of the electric charge, and to a good
approximation is independent from the particle energy and direction.
We speculate that this deterministic energy loss 
is an important, or perhaps even the 
leading contribution to the solar modulation effects.
\end{abstract}

\section{Introduction}
\label{sec:introduction}
The fluxes of galactic cosmic rays observed near the Earth 
are different from those present in the 
local interstellar space because the particles lose energy
and are deviated when they penetrate the heliosphere.
These effects are time dependent because the heliospheric environment
is not static. The time variations have
several different characteristic time scales with a prominent
22 years periodicity associated to the 
solar magnetic activity. The solar modulations 
of the cosmic rays energy spectra 
have been extensively studied experimentally for many years
using ground and space based detectors,
and has been the subject of many theoretical studies
\cite{Potgieter:2013pdj}.

Following the pioneering work of Parker \cite{Parker}, 
the usual way to study theoretically the solar modulations
is to model the propagation of cosmic rays in the heliosphere
as a diffusion process in a turbulent magnetic field,
accompanied by convection (because of the motions of the plasma
-- the solar wind -- that fills the heliosphere) 
and adiabatic energy losses
associated to the expansion of the plasma.
The effects of diffusion, convection and adiabatic energy loss
in particle propagation are included in the Parker equation 
that has been used as the standard instrument to 
study solar modulations
(for a recent reviews see \cite{Strauss:2012zza}).

In the first studies of the solar modulations 
(such as \cite{Goldstein:1970us})
the calculated effects were identical for particles of 
opposite electric charge.
The data however show that the modulations effects differ
for particles of opposite electric charge. 
The dependence of the modulations on the sign of the electric charge 
can be attributed to the presence of a regular magnetic field 
in the heliosphere.
In the presence of such a field, the average trajectories of particles 
with opposite charge differ, and the particles
can have different energy losses.

The effects of the regular heliospheric field 
was first included in the Parker equation by Jokipii \cite{Jokipii-1977}
introducing an additional convection term 
associated to the average ``drift velocity'' of the gyration center
of charged particles moving in a non--homogeneous magnetic field.
The direction of the drift velocity is reversed when the electric charge
of a particle changes sign, and therefore the drift velocity 
convection term breaks the particle/antiparticle symmetry
in the Parker equation \cite{Bobik:2011ig,Maccione:2012cu,Potgieter:2014pka}.

In this work we will discuss the propagation of charged particles 
in the heliosphere assuming that only the regular electromagnetic field is present, 
neglecting completely the effects of the random component of the field.
This approach is complementary to the traditional method
(where it is the regular field that is completely neglected), and a comparison
of the results can be instructive to clarify the physical mechanisms
that generate the solar modulations.

To keep the discussion as simple as possible we will assume 
that the heliospheric magnetic field $\vec{\Bf}(\vec{x})$ 
has the simple Parker spiral form (discussed below), and 
that the speed of the solar wind is constant, and its 
direction is radial ($\vec{w}(\vec{x}) = w \, \hat{r}$).
The magnetic field and the solar wind
will be also considered as stationary during the time necessary 
for charged particles to reach the Earth from
the boundary of the heliosphere.

If the conductivity of the plasma in the heliosphere 
is sufficiently high, the condition of stationarity 
implies that the net force on an electrically charged particle moving 
with the wind must vanish. It follows that 
in a frame where the Sun is at rest there is a non vanishing 
electric field given by the well known expression:
\begin{equation}
\vec{\Ef}(\vec{x}) = -\frac{\vec{w} (\vec{x})}{c} \wedge \vec{\Bf}(\vec{x}) ~.
\label{eq:e-field}
\end{equation}
Equation (\ref{eq:e-field}) can be also derived 
requiring that the electric field vanishes
in a frame where the plasma of the solar wind is at rest.

Under the assumptions we have outlined above, the propagation 
of a charged particle in the heliosphere, is completely
deterministic and can be calculated integrating the classical
equations of motion:
\begin{eqnarray}
\frac{d\vec{x}} {dt} 
= \vec{v} = \frac{\vec{p}} {\sqrt{p^2 + m^2}} 
\nonumber \\
& ~ & \label{eq:dynamics} \\
\frac{d\vec{p}} {dt} = 
q\; \left ( \vec{\Ef} + \frac{\vec{v}}{c} \wedge \vec{\Bf} \right ) \nonumber ~.
\end{eqnarray}

In the following we will show that charged particles of sufficiently low energy
($E \lesssim 1$~TeV) 
propagating from the 
boundary of the heliosphere to the Earth will always {\em lose} energy,
and discuss the properties and consequences of this energy loss.


This paper is organized as follows:
in the next section we describe the model of the 
heliospheric electromagnetic fields used in this work.
In section~\ref{sec:drift} we discuss the magnetic drift velocities
in the motion of charged particles in our heliospheric model.
In section~\ref{sec:trajectories} we discuss some examples of 
trajectories of charged particles in the heliosphere.
Section~\ref{sec:eloss} discusses the energy loss suffered by particles
that arrive at the Earth. Section~\ref{sec:energy-spectra} discusses the relation between 
the interstellar spectra of cosmic rays at the boundary of the heliosphere,
and the energy spectra observable at the Earth.
The last section gives a summary and outlook.

\section{Electromagnetic fields in the heliosphere}
\label{sec:fields}
\begin{figure}
\begin{center}
\includegraphics[width=12.0cm]{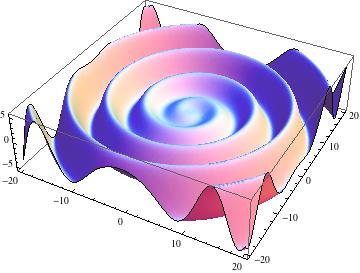}
\end{center}
\caption {\footnotesize
Representation of the heliospheric current sheet for tilt angle $\sin\alpha = 0.2$. 
\label{fig:ballerina1}
 }
\end{figure}

\begin{figure} [ht]
\begin{center}
\includegraphics[width=12.0cm]{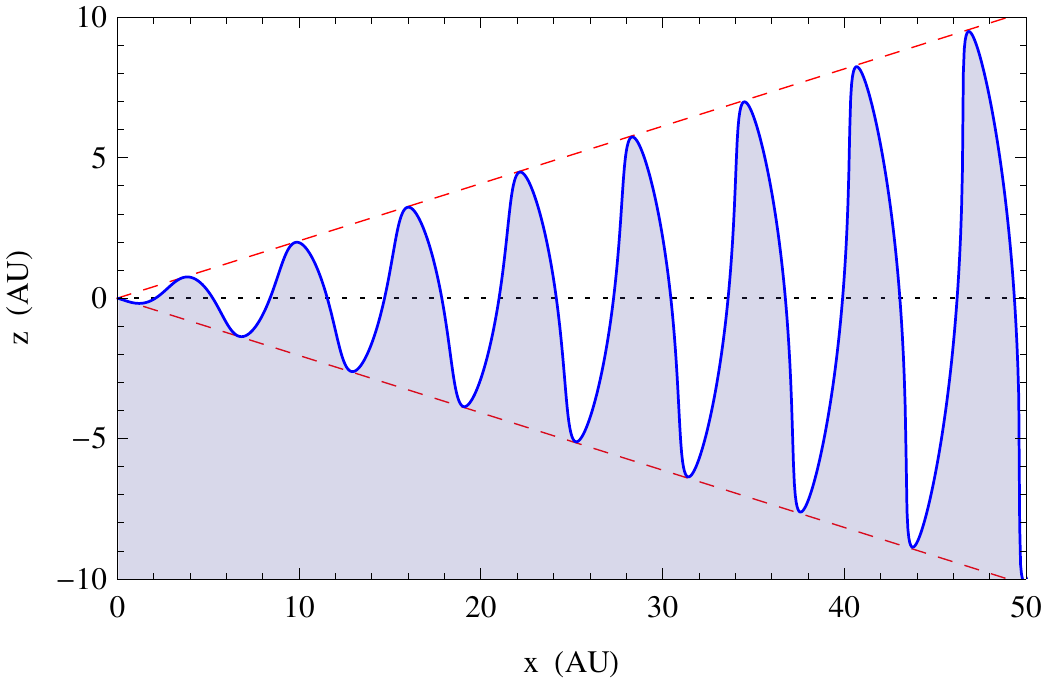}
\end{center}
\caption {\footnotesize
Section $x z$ of the heliospheric current sheet. 
For positive polarity ($A >0$)
the field lines enter (exit) the plane of the figure in
the white (shaded) area.
The dashed lines correspond to the
condition $\cos \theta = \pm \sin \alpha$ (with $\theta$ is the 
standard the polar angle and $\alpha$ the tilt angle of the heliospheric
field).
For a negative polarity phase the field directions are inverted.
\label{fig:sheet_section}
 }
\end{figure}

In the following we will assume that the 
solar magnetic field has the Parker spiral form:
\begin{equation}
\vec{\Bf} = A~B_0 \; \left ( \frac{r_0}{r} \right )^2
\; \left [ \hat{r} - \frac{\Omega}{w} \, r \, \sin\theta 
\; \hat{\varphi} \right ] \; S(\vec{r})
\label{eq:parker-field}
\end{equation}
In this equation $r$, $\theta$ and $\varphi$ are spherical coordinates
in a frame centered on the Sun, 
$\hat{r}$ and $\hat{\varphi}$ are versors, 
 $r_0$ is the radius of the Earth orbit (approximated as a circle),
$\Omega$ is the (equatorial) frequency of the solar rotation, 
$w$ is the speed of the solar wind that we will assume is radial and 
constant ($\vec{w} = w \; \hat{r}$), 
$B_0$ is a normalization constant, 
$A=\pm 1$ give the polarity of the solar cycle
(that is reversed approximately every 11 years),
and $S(\vec{r})$ is $\pm 1$:
\begin{equation}
S(\vec{r}) = \begin{cases} 
+1 & \mbox {if } \cos\theta > \cos \theta^* \\
-1 & \mbox {if } \cos\theta \le \cos \theta^* 
\end{cases}
\end{equation}
with 
\begin{equation}
\cos \theta^* = -\sin \alpha \,
\sin \left (\varphi + \frac{\Omega}{w}\, r \right ) ~.
\label{eq:sheet}
\end{equation}
The set of points that satisfy the equation $\cos\theta = \cos\theta^*$
forms a surface known as the Heliospheric
Current Sheet (HCS) where the magnetic field is
discontinuous. 
The ``tilt angle'' $\alpha$ can take values in the interval $[0,\pi/2]$.
For $\alpha=0$ the HCS coincides with the $xy$ equatorial plane,
more in general the surface has the wavy form of a ``ballerina skirt''.
The shape of the current sheet is illustrated 
in fig.\ref{fig:ballerina1}, \ref{fig:sheet_section} and~\ref{fig:plane}.


It is convenient to introduce the characteristic length:
\begin{equation}
R = \frac{w}{\Omega} \simeq 0.93~\left [\frac{w}{400 ~{\rm km/s}} 
\right ]{\rm AU} \;,
\label{eq:rspir}
\end{equation}
that will enter in several expressions in the following.
The modulus of the magnetic field is:
\begin{equation}
\left |\Bf \right (\vec{x}) | = B_0 \; \left ( \frac{r_0}{r} \right )^2
\sqrt{1 + \frac{\Omega^2 }{w^2} \; (r^2-z^2)}~.
\end{equation}
When $r$ increases for a fixed direction $\hat{n}$,
the field decreases in general  $\propto r^{-1}$, with the only exception 
of the $\pm \hat{z}$ directions, when the field is $\propto r^{-2}$.
The field at the Earth is:
\begin{equation}
B_\oplus = B_0 \;
\sqrt{1 + \left (\frac{\Omega \, r_0}{w} \right )^2 }~.
\end{equation}

The magnetic field lines have the form of clockwise Archimedes spirals
that are enveloped on a cone with vertical axis and vertex at the 
origin of the coordinates.
The field line that passes through the point with
polar coordinates $\{\overline{r}, \overline{\theta}, \overline{\varphi} \}$ 
can be parametrized in terms of the azimuth angle $\varphi$ of the points along the line:
\begin{eqnarray}
\theta (\varphi) & = & \overline{\theta} \nonumber \\
& ~ & \label{eq:line1} \\
r (\varphi) & = & 
\overline{r} + R \times \left [ \overline{\varphi} - \varphi \right ]
\nonumber 
\end{eqnarray}
with the angle $\varphi$ in the interval:
\begin{equation}
-\infty < \varphi \le \varphi_{\rm max} = \overline{\varphi} + \frac{\overline{r}}{R}~.
\label{eq:line2}
\end{equation}
The range of variation of $\varphi$ in the previous equations
shows that each field line extends to infinity in one direction and reaches
the origin at $r=0$ after a finite length in the opposite direction.
The distance from the origin of a point on the line 
grows linearly with $\varphi_{\rm max} - \varphi$, and the 
quantity $\varphi_{\rm max}$ gives the azimuth angle of the 
field line direction at the origin.

For positive polarity ($A > 0$)
the magnetic lines are outward (inward) in the hemisphere
 above (below) the heliospheric current sheet.
For negative polarity 
 ($A < 0$) the direction of all field lines is reversed.

The heliospheric electric field can be calculated using equation 
(\ref{eq:e-field}). In cartesian components one has:
\begin{equation}
\vec{\Ef} (x,y,z) = 
\pm A \, B_0 \;
\frac{\Omega \, r_0^2}{c \, r^3} \left \{ x \, z, y\, z, -(x^2 + y^2) \right \}
\end{equation}
where the $+$ ($-$) sign applies above (below)
the heliospheric current sheet. Note that the near the heliospheric equator
the electric field is vertical, and points toward the HCS for $A > 0$, and 
away from the HCS for $A <0$.
The streamlines of the electric field are shown in fig.\ref{fig:electric-field}.
The modulus of the electric field is 
\begin{equation}
\left |
\vec{\Ef} (x,y,z) 
\right |
= B_0 \; \frac{\Omega \, r_0^2}{c} 
\; \frac{\sqrt{x^2 + y^2}}{r^2}
\end{equation}
The divergence and rotor of $\vec{\Ef}$ are: 
\begin{equation}
\nabla \cdot \vec{\Ef} = 
\pm A\, B_0 \; \frac{\Omega \, r_0^2}{c} \; \frac{2 z}{r^3}
\label{eq:divergence}
\end{equation}
and
\begin{equation}
\nabla \wedge \vec{\Ef} = 
0~.
\label{eq:rotor}
\end{equation}
Equation (\ref{eq:divergence}) implies 
 that a non vanishing electric charge density must be present in the 
heliosphere to satisfy the Maxwell equations.
Equation (\ref{eq:rotor}) has very important consequences, because it 
implies that the electric field can be written as the gradient of a potential function:
\begin{equation}
\vec{\Ef} (x,y,z) = -\nabla V(x,y,z)~.
\label{eq:gradient}
\end{equation}
Choosing the constant so that the potential vanishes at the 
heliospheric equator $z=0$, the electric potential $V$ is:
\begin{equation}
V(x,y,z) = 
\mp A \, B_0 \;
\frac{\Omega \, r_0^2}{c }\; \frac{z}{r} ~
\label{eq:vpotential}
\end{equation}
where the $\mp$ sign is valid above (below) the heliospheric current sheet.
Note that on the HCS the electric potential $V$ (as also the electric and magnetic fields)
is discontinuous for all points that have $z \ne 0$. 
A $z=0$, when the HCS coincides with the equatorial plane,
the  potential vanishes and is continuous
(but the electric and magnetic fields are discontinuous).

The limits of the potential for $z \pm \infty$ 
(with $x$ and $y$ finite) are equal to each other, and independent from
the values of $x$ and $y$: 
\begin{equation}
\lim_{z \pm \infty} V(x,y,z) = 
V_\infty =
-A \, \left |V_\infty \right | =
- A 
\, B_0 \; 
\frac{\Omega \, r_0^2}{c} ~.
\label{eq:vinfty}
\end{equation}
Note that  $V_\infty$, 
the asymptotic value  of the potential for $|z| \to \infty$,
changes sign with the polarity of  the solar phase
(is negative for $A > 0$, and positive for $A < 0$).
The existence of the electric field potential and the result that
all points with large $|z|$ have the same potential have
important consequences that will be discussed later.

It should be noted that the Parker spiral expression of
equation (\ref{eq:parker-field}) 
cannot be a good description of the magnetic field in all space,
and it must fail both for very small and very large distances
from the origin of the coordinates. 
This can be verified computing the total energy of the magnetic field.
Integrating the energy density $B^2/(8 \pi)$ of the Parker spiral field
in the volume contained between 
radii $r_{\rm min}$ and $r_{\rm max}$ one obtains:
\begin{equation}
E_{\rm magnetic} = \frac{B_0^2 \, r_0^4}{2} 
\; \left [
\frac{1}{r_{\rm min}}
-\frac{1}{r_{\rm max}} +
\frac{2}{3} \, \frac{(r_{\rm max}-r_{\rm min})}{(w/\Omega)^2}
\right ]~.
\end{equation}
This expression diverges in the limits 
$r_{\rm min} \to 0$ and $r_{\rm max} \to \infty$ demonstrating that 
a Parker spiral magnetic field that extends to all space 
is unphysical.
The extension of the Parker spiral form to $r \to \infty$
would in fact correspond to the assumption of a
solar wind emitted with constant velocity 
for an infinitely long time in the past.

The expression (\ref{eq:parker-field}) for the heliospheric magnetic field
should therefore be considered as approximately valid only in a finite volume
outside an inner spherical surface of few solar radii,
and inside the termination shock, a roughly spherical surface 
with a radius of approximately 100~AU 
where the solar wind suffers an abrupt deceleration.

In this work we will not attempt to model the magnetic field
outside the termination shock, that is in fact only very poorly known.
The results shown here are not very sensitive to
the size of the volume where the field is well described as a Parker spiral,
if it extends to a radius equal or larger than $\sim 100$~AU.

\section{Drift velocity}
\label{sec:drift}
The motion of a particle in a non homogeneous magnetic field
can be decomposed into the gyration around the field lines
accompanied by the drift of the gyration center in 
a direction orthogonal to the lines \cite{jackson}.
The drift velocity is given by the well known expression:
\begin{equation}
\vec{v}_{\rm drift} = 
\frac{ \vec{\Bf} \wedge \vec{\nabla} |\Bf|}{2 \,q \, |\Bf|^3}
\; E \, c\; \left (\beta_\perp^2 + 2 \, \beta_\parallel^2 \right )
\label{eq:vdrift}
\end{equation}
where $q$ and $E$ are the charge and energy of
the particle, and $\beta_\perp$ and $\beta_{\parallel}$ are 
the components of the particle velocity orthogonal and
parallel to the magnetic field.
The two terms proportional to $\beta_\perp^2$ and
$\beta_\parallel^2$ are known as the gradient and curvature contributions
to the drift velocity.

For a magnetic field of the Parker spiral form the drift velocity can
be calculated explicitely as: 
\begin{equation}
\vec{v}_{\rm drift} = \pm A \,\frac{q}{|q|} \, V_\beta\, 
\; \left [
\hat{\rperp} \; \frac{2 \, \rperp \, z}{R^2 + z^2}
+\hat{\varphi} 
\; \frac{R \rperp z \, \sqrt{\rperp^2 + z^2}}{(R^2 + z^2)^2}
-\hat{z} 
\; \frac{\rperp^2\, (2 R^2 + \rperp^2 - z^2)}{(R^2 + z^2)^2}
\right]
\end{equation}
where again the $\pm $ sign applies
above (below) the heliospheric current sheet,
$\hat{\rperp}$, $\hat{\varphi}$ and $\hat{z}$ are versors
in the three directions of cylindrical coordinates 
(with $\rperp = \sqrt{x^2 + y^2}$), the length $R$ was
defined in equation (\ref{eq:rspir})
and $V_\beta$, with the dimensions of a velocity, is:
\begin{equation}
V_\beta = V_0~ \left (\beta_\perp^2 + 2 \, \beta_{\parallel}^2 \right ) 
\end{equation}
with 
\begin{equation}
V_0 = \frac{E}{2 |q| \,B_0}\, \frac{w}{\Omega} \, \frac{c}{r_0^2} 
\simeq c ~\left (1.04 \times 10^{-3} \right )
\, \left [\frac{E}{{\rm GeV}} \right ]
\, \left [\frac{10^{-4}\, {\rm G}}{B_0} \right ]
\, \left [\frac{w}{400\;{\rm km/s}} \right ]
\end{equation}
The direction of the drift velocity 
is inverted when the electric charge of the particle 
changes sign, and also when the polarity of the solar phase is reversed.
In the equatorial plane ($z=0$) the drift velocity
is parallel to the $\pm \hat{z}$ direction and has the simple expression:
\begin{equation}
\vec{v}_{\rm drift}(z=0) = \mp A ~V_\beta 
\left [ 1 - \frac{R^4}{(R^2 + \rperp^2)^2} \right ]~\hat{z} \;.
\end{equation}
The term in square parenthesis vanishes for $\rperp =0$ and 
grows monotonically with $\rho$, reaching an asymptotic value of 
one when $\rho \to \infty$. 
For $qA > 0$ (positive solar polarity and positive electric charge,
or negative polarity and negative charge)
the $z$ component of $\vec{v}_{\rm drift}$ 
is negative above the HCS, and positive below the HCS,
in other words the drift velocity transports the particles toward the current sheet.
In the opposite case ($q A <0$) the effect is reversed and 
the drift transports the particles away from the current sheet.

The streamlines of the drift velocity are shown in fig.\ref{fig:vdrift}.
Comparing these figure with the streamlines of the electric field
shown in fig.\ref{fig:electric-field} one can notice that the directions
of the drift velocity and of the electric force
appear to be always antiparallel.
In fact the scalar product of the drift velocity and 
the electrical force can be calculated explicitely
with a result:
\begin{equation}
\left \langle 
\frac{dE}{dt}
\right \rangle =
q \, \vec{\Ef}\cdot \vec{v}_{\rm drift} =
-\frac{|q|}{2} E\, w \;
\frac{\rperp^2 (2 R^2 + \rperp^2)}{(R^2 + \rperp^2)^2 \, r}
\; \left (\beta_\perp^2 + 2 \, \beta_{\parallel}^2 \right )
\label{eq:elossav}
\end{equation}
that is manifestly negative.
This result is valid for any sign of the 
electric charge and of the solar polarity, 
because reversing the sign of $q$ or $A$ 
inverts the directions of both the electrical force 
and of the drift velocity,
leaving the scalar product of the two vectors invariant.
The scalar product $q \, \vec{\Ef} \cdot \vec{v}_{\rm drift}$,
as indicated in equation (\ref{eq:elossav}),
is equal to the energy variation of a charged particle that moves in the heliosphere
on a trajectory controled by the electromagnetic field (after averaging over
the fast gyration around the magnetic field lines).
Equation (\ref{eq:elossav}) therefore states that the average energy variation is
always negative, it is an energy {\em loss}.

It is instructive to compare equation (\ref{eq:elossav})
with the expression of the adiabatic energy loss
used in the Parker equation for 
a relativistic particle diffusing in a spherically expanding wind:
\begin{equation}
-\left \langle 
\frac{dE}{dt}
\right \rangle = |q| \, E \; \vec{\nabla} \cdot \vec{w} = \frac{2 |q| \, E \, w}{r} 
~.
\label{eq:eloss-parker}
\end{equation}
The expressions (\ref{eq:elossav}) and (\ref{eq:eloss-parker})
are very similar, with identical
dependences ($\propto |q| \,E \, w$) on the wind velocity
and on the charge and energy of the particle, and similar dependence
on the space coordinates (approximately $\propto r^{-1}$ in both cases).
This  similarity emerges because  the expression of the adiabatic energy loss
can be derived  with magnetohydrodynamics consideration 
for the random magnetic field that are very similar to those 
developed here  for the large scale regular field.
The important difference between
expression (\ref{eq:elossav}) and (\ref{eq:eloss-parker}) is that the first case
the energy loss  can be associated to an electric  potential, while in the other
this is not possible.  This point will be illustrated more extensively
in section~\ref{sec:eloss}.

\section{Space trajectories in the heliosphere}
\label{sec:trajectories}
\begin{figure} [ht]
\begin{center}
\includegraphics[width=12.0cm]{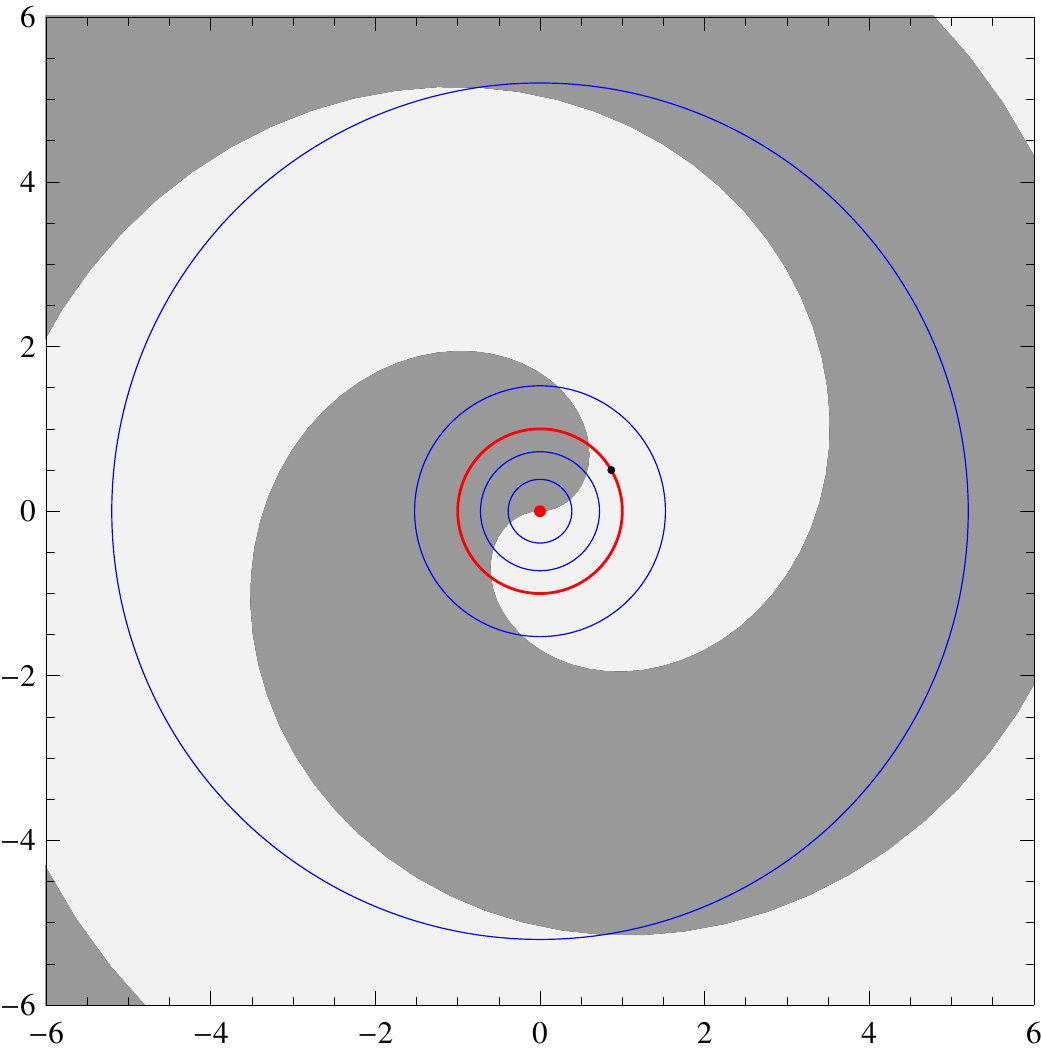}
\end{center}
\caption {\footnotesize Section $x y$ of the
inner part of the heliosphere. The Sun is at the center of the image,
the circles indicate the orbits of the five inner planets 
(Mercury, Venus, Earth, Mars and Jupiter).
The spiral that separates the darker and lighter regions
marks the intersection of the heliospheric current sheet with
the Earth's orbital plane. 
In the darker (lighter) region the current sheet is 
below (above) the plane.
In a positive polarity solar phase the magnetic field lines are 
directed toward (away from) the Sun below (above) the current sheet.
For a negative polarity phase the field directions are inverted.
\label{fig:plane}
}
\end{figure}

\begin{figure} [ht]
\begin{center}
\includegraphics[width=12.0cm]{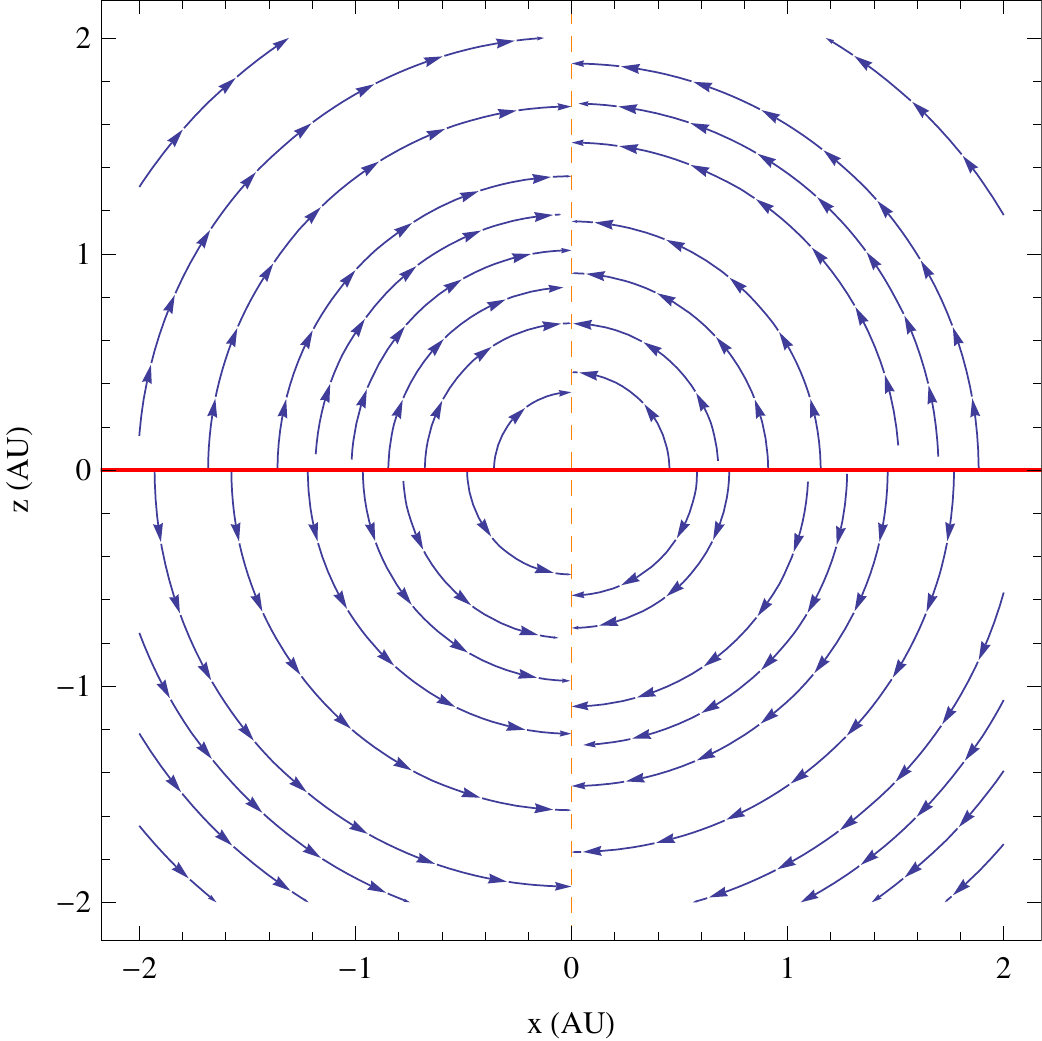}
\end{center}
\caption {\footnotesize 
Streamlines of the electric field in the heliosphere.
The thick line represents the heliospheric current sheet.
The electric field $\vec{Ef}$ 
has cylindrical symmetry and vanishes along the line $x = 0$.
The figure is valid for $A > 0$.
 For $A <0$ the field direction
is reversed. 
\label{fig:electric-field}
}
\end{figure}

\begin{figure} [ht]
\begin{center}
\includegraphics[width=12.0cm]{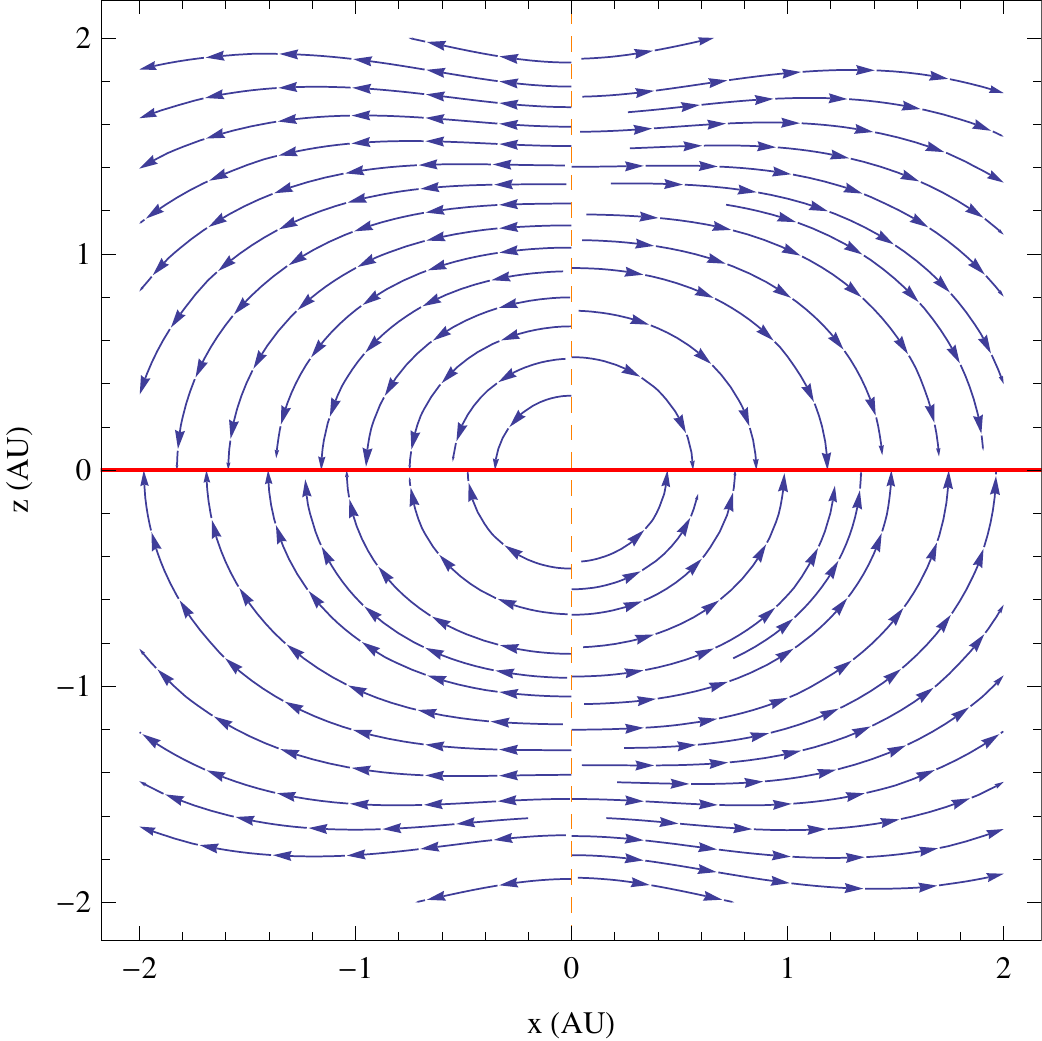}
\end{center}
\caption {\footnotesize 
Projection in the plane $(x,z)$
of the streamlines of the drift velocity 
of charged particles moving in the heliosphere. 
The figure is valid for $q A > 0$. The direction
Changing the sign of the electric charge or the polarity
of the solar phase the direction of the drift is reversed.
\label{fig:vdrift}
}
\end{figure}

\begin{figure} [ht]
\begin{center}
\includegraphics[width=14cm]{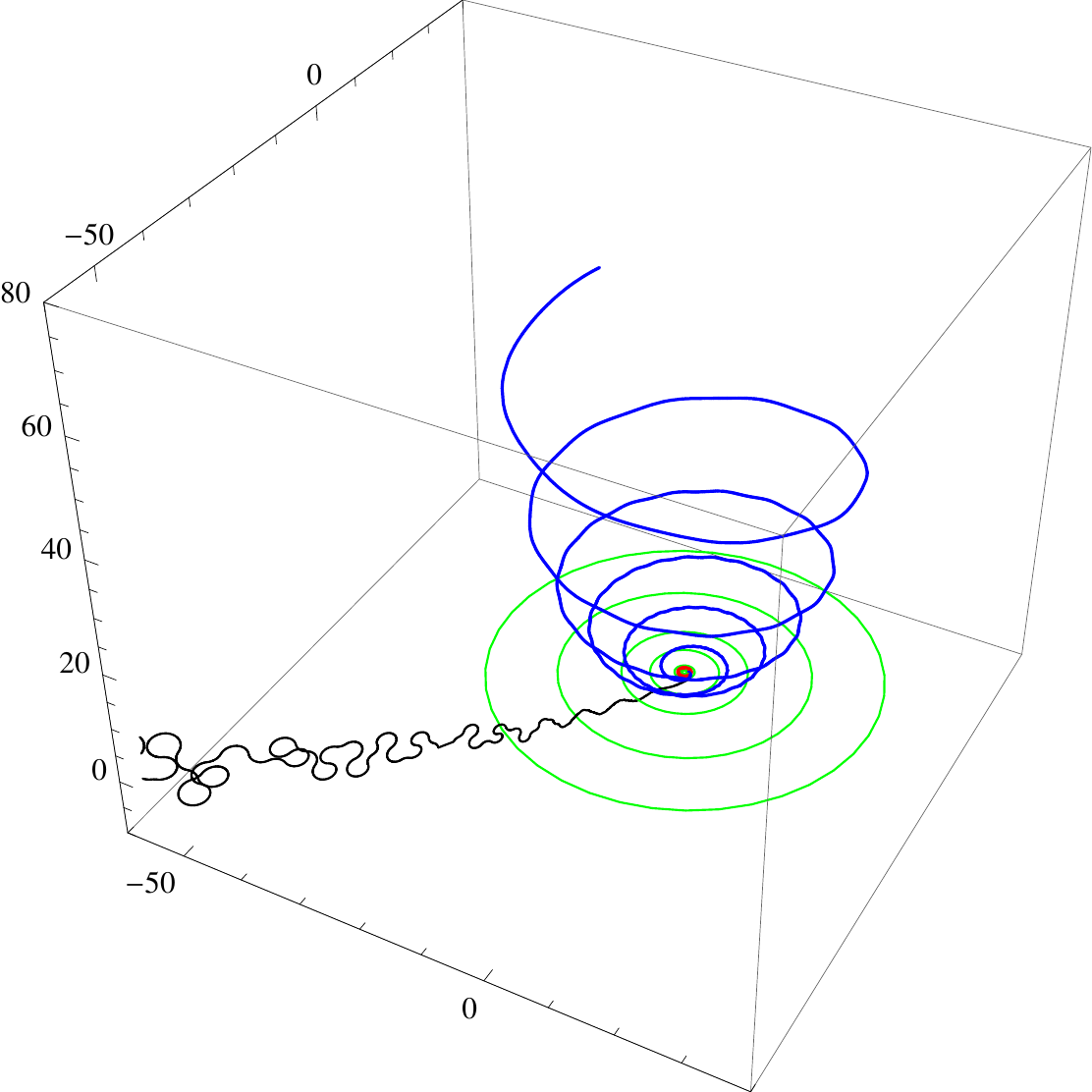}
\end{center}

\caption {\footnotesize
Space Trajectory of an electron in the heliosphere. The particle is 
observed at the Earth with energy 10~GeV. 
The thick (thin) line shows the trajectory before (after) the observation.
The green circles are the trajectories of the planets.
\label{fig:example3D}
 }
\end{figure}

For a qualitative understanding of the problem we are discussing,
it is very instructive to inspect some examples of 
trajectories of charged particles in the heliosphere,
obtained with a numerical integration of 
the equation of motions (\ref{eq:dynamics}).
In the  following  calculations we will use 
a coordinate system where the Sun is at the origin and the $z$ axis 
is parallel to the Sun rotation axis; the orientation of the
$x$ and $y$ axis with respect to the pattern of the HCS is  chosen as shown
in fig.\ref{fig:plane}.
One example of trajectory is shown in fig.\ref{fig:example3D} 
for an electron that is ``observed'' at the Earth with energy $E=10$~GeV
and direction with polar angles
$\{\cos \theta = 2/3, \varphi = 3 \pi/4\}$.
At the ``detection'' time the phase of the Earth with respect to the
pattern of the heliospheric current sheet
was $\varphi_\oplus = 0$, when the Earth 
(see fig.\ref{fig:plane}) is above the heliospheric current sheet.
The trajectory is calculated from this initial condition
integrating the equations of motion both forward and backward in time.
In the calculation the solar polarity is positive ($A > 0$) and the
parameters that describe the Sun magnetic field 
are $B_0 = 10^{-4}$~Gauss, $w = 400$~km/s and $\sin\alpha =0.05$. 
Reversing simultaneously the polarity $A$ of the heliospheric 
magnetic field and the electric charge $q$ of the particle 
the trajectory remains identical and therefore
the example of fig.\ref{fig:example3D} 
refers more in general to the condition $q A > 0$. 

The particle of this example reaches the Earth,
traveling on a down--going spiral 
that wraps around a surface of approximately paraboloidal shape, 
and then leaves the heliosphere traveling close to the 
equatorial plane, or more accurately close to the heliospheric current sheet.

\begin{figure} [ht]
\begin{center}
\includegraphics[width=14cm]{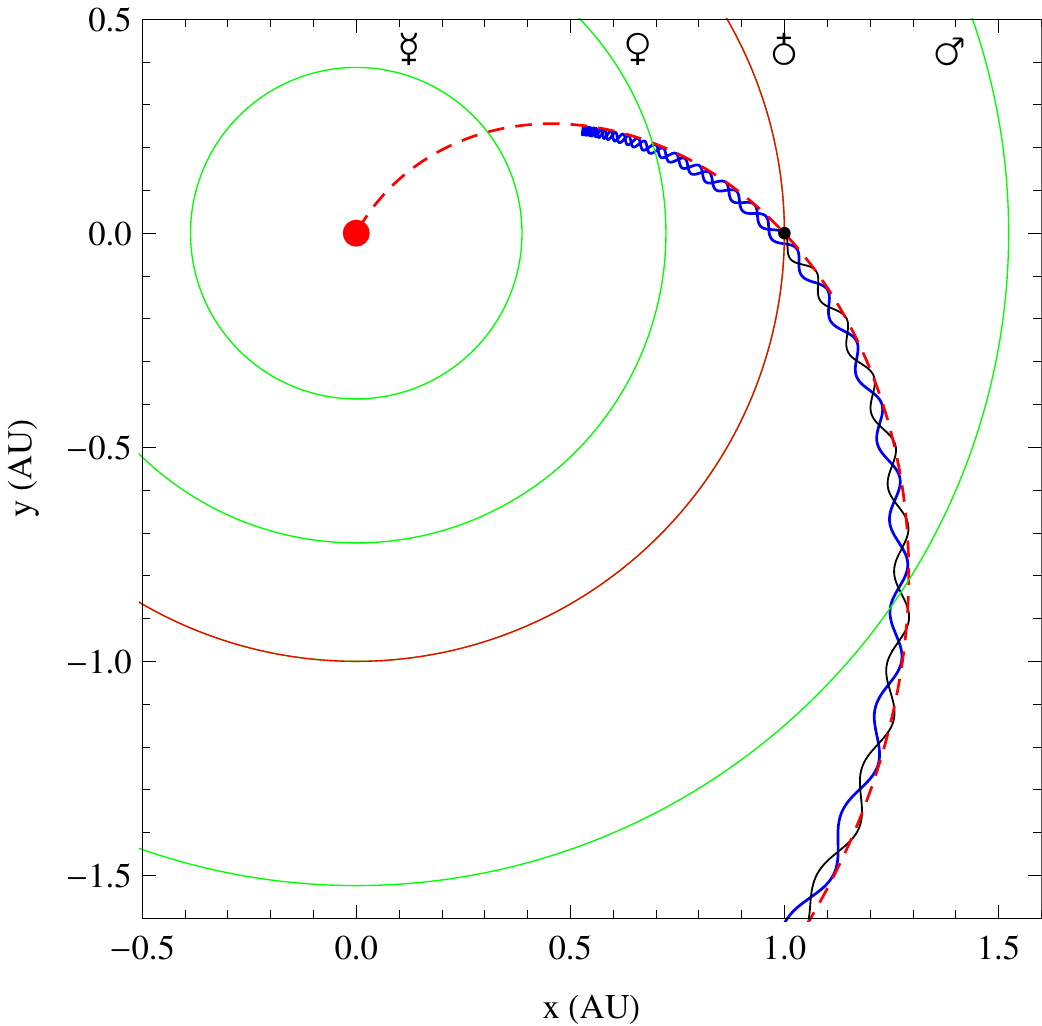}
\end{center}

\caption {\footnotesize
Projection in the $\{x,y\}$ plane of the space
trajectory shown in fig.\ref{fig:example3D}. 
The thick (thin) line corresponds the portion of the trajectory
before (after) the observation at the Earth.
The dashed red line shows the magnetic field line that 
passes through the observation point. 
\label{fig:ex1}
 }
\end{figure}

\begin{figure} [ht]
\begin{center}
\includegraphics[width=14cm]{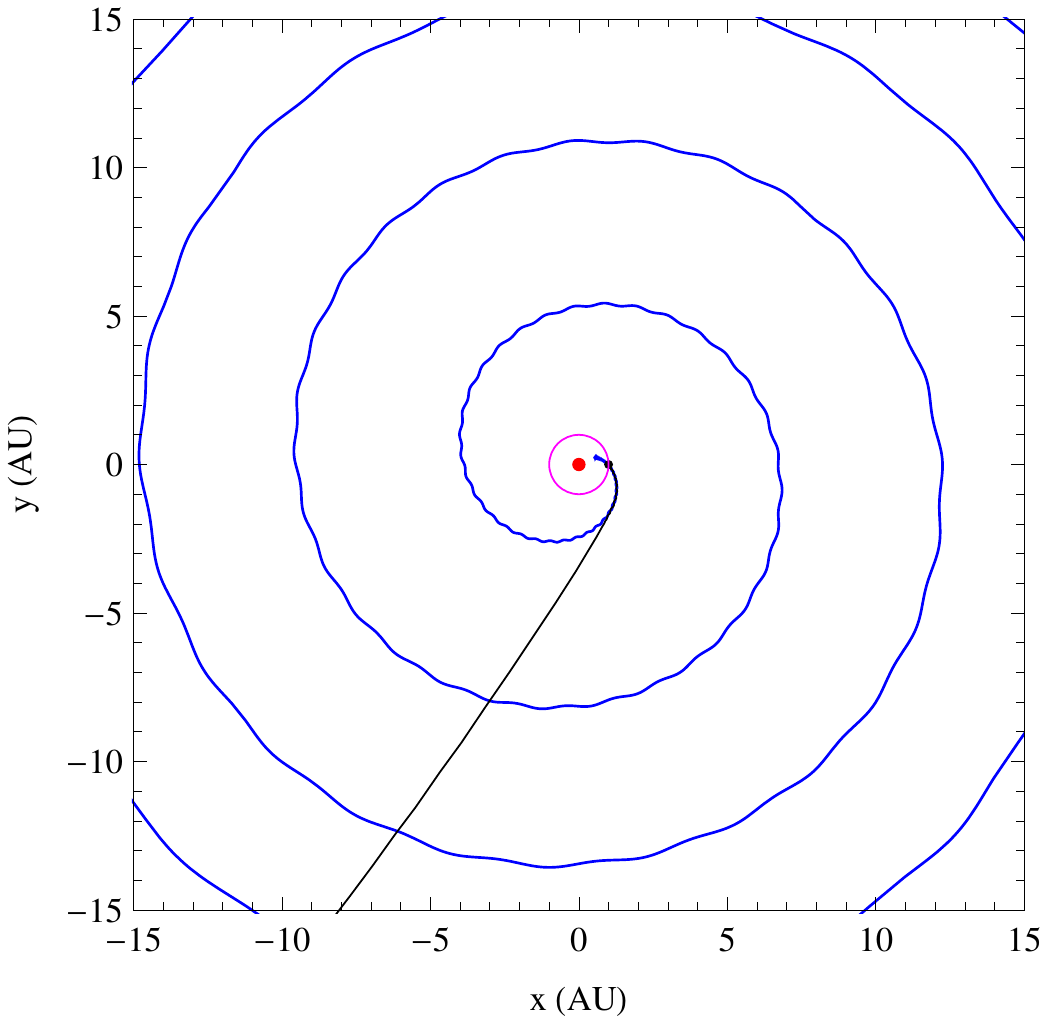}
\end{center}

\caption {\footnotesize
Same as fig.\ref{fig:ex1} but with an enlarged scale.
\label{fig:ex2}
 }
\end{figure}

\begin{figure} [ht]
\begin{center}
\includegraphics[width=14cm]{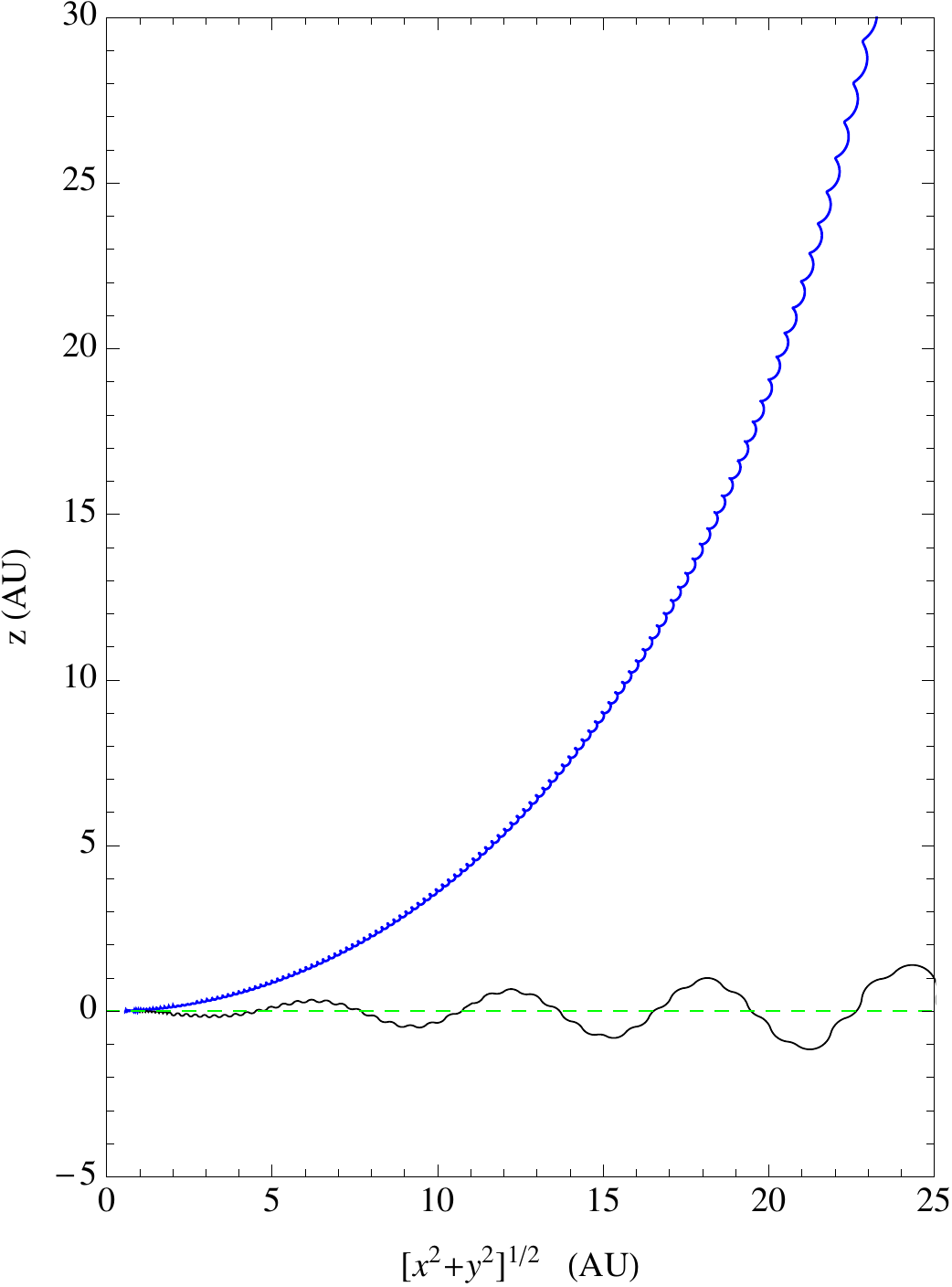}
\end{center}

\caption {\footnotesize
Projection in the $\{\sqrt{x^2+y^2},z\}$ space 
of the trajectory shown in fig.\ref{fig:example3D}
(a particle with observed energy 10~GeV).
The thick (thin) line corresponds the portion of the trajectory
before (after) the observation at the Earth.
\label{fig:ex3}
 }
\end{figure}

For a better visualization,
fig.\ref{fig:ex1} and fig.\ref{fig:ex2} show (with different scales)
the projections of the trajectory in the equatorial \{$x,y$\} plane, while 
fig.\ref{fig:ex3} shows the projection 
in the $\{\sqrt{x^2 + y^2}, z\}$ space.
In fig.\ref{fig:ex1}
one can best see (in the $\{x,y\}$ projection) 
the trajectory in the inner part of the heliosphere.
It is clear that, in first approximation, the particle 
gyrates around a magnetic field line 
(one such line is shown as a dashed curve in the figure). The trajectory also
undergoes a ``bounce'', (or a ``reflection'')
at the point of closest approach to the Sun, an example 
of the well known magnetic mirror effect, present 
when the magnetic field lines converge.
Fig.\ref{fig:ex2} and fig.\ref{fig:ex3} allow to visualize 
the ``in'' part of the particle trajectory as a spiral wrapped around a paraboloid, 
and the ``out'' part as propagation close to the wavy heliospheric current sheet.

The main qualitative features of the trajectory we have just presented:
the spiral, the bounce, and the outgoing travel close to the HCS, are in fact
general features that 
are present in all trajectories when the product of the particle electric
charge and of the solar polarity is positive ($q A > 0$).
The spiral is downgoing (as in our example) when the Earth is located above the HCS,
but it is upgoing when the Earth is below the HCS.
In the opposite case ($q A < 0$) 
the trajectories have the same three features, but in reversed order.
The particles enter the heliosphere 
traveling close to HCS, undergo a bounce
at a point of closest approach to the Sun, and then exit the heliosphere
traveling on a (upgoing or downgoing) spiral.

The qualitative features of the trajectories, 
are determined by the structure of magnetic field, 
(note that in the equations of motion 
the effects of the electric field are suppressed by a factor $w/c$, and can be 
considered as a as small perturbation), and are easy to understand qualitatively,
using the results on the properties of the magnetic field lines and of the
drfit velocity derived in the previous sections.
The key points are the facts that the magnetic field lines
are spirals around vertical cones, and that the drift velocity points 
toward (away from) the current sheet for $q A >0$ ($q A < 0$).

For $q A >0$, when the drift ``pushes'' the particle toward the HCS,
the charged particle can enter the heliosphere only 
traveling along a trajectory that starts at large $|z|$,
and spirals around the $z$ axis following the behavior of the field lines.
Since all magnetic field lines converge at the origin, 
and the convergence acts as a magnetic mirror,
the particle trajectory will stop approaching the Sun, 
and will be reflected outward, undergoing a ``bounce''.
The qualitative behavior of the trajectory changes dramatically when
the particle reaches the HCS. The magnetic drift effects
(for the case $q A > 0$ we are discussing now)
always push the particle toward the HCS, 
so that it remains ``trapped'' close to the sheet
(crossing it multiple times) while it travels out of the heliosphere.

A similar discussion can be applied to the case
$q A < 0$ to show that the trajectory is formed by 
the sequence of an ``in'' section, where the particle
reaches the Earth traveling near the HCS, a bounce, and an ``out'' section
that is an (upgoing or dowgoing) spiral around the $z$ axis.

\begin{figure} [ht]
\begin{center}
\includegraphics[width=4.4cm]{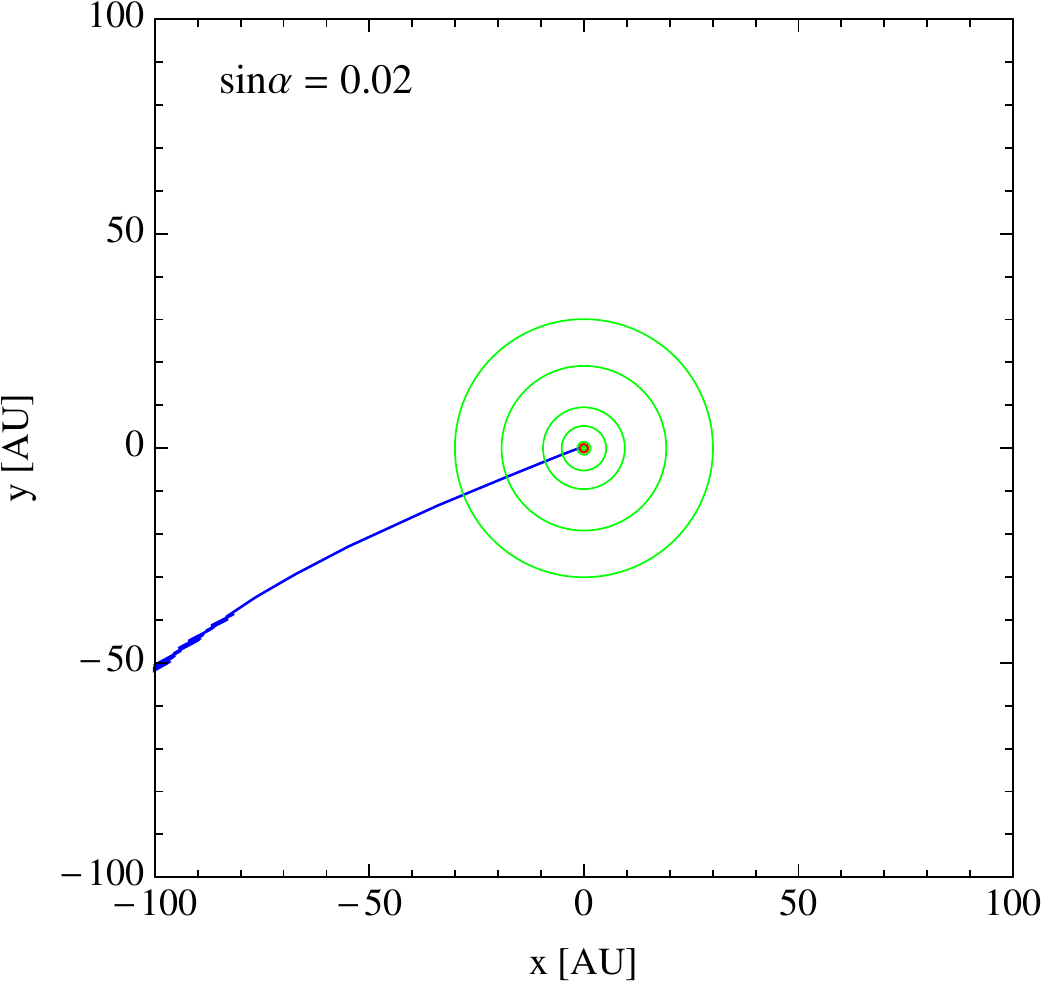}
~~~~\includegraphics[width=4.2cm]{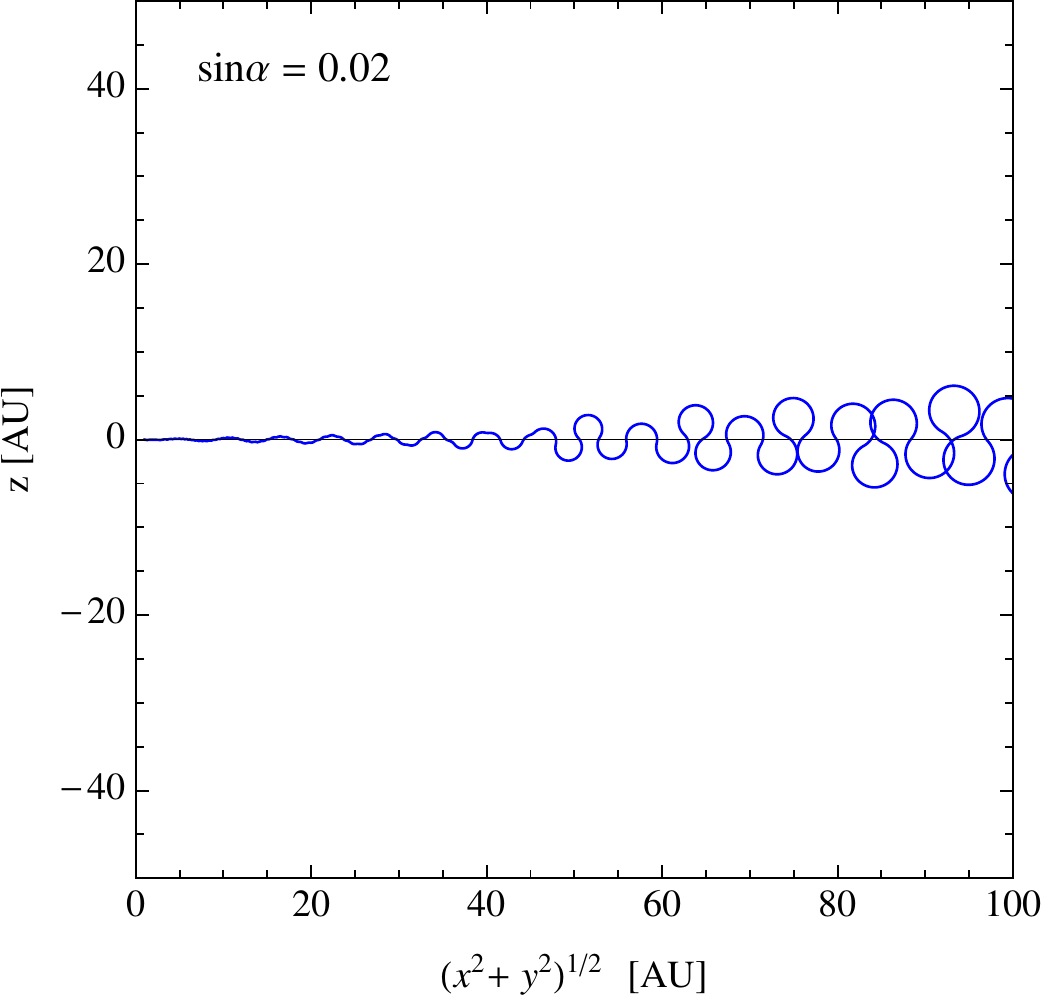}
\end{center}

\vspace{0.2cm}
\begin{center}
\includegraphics[width=4.4cm]{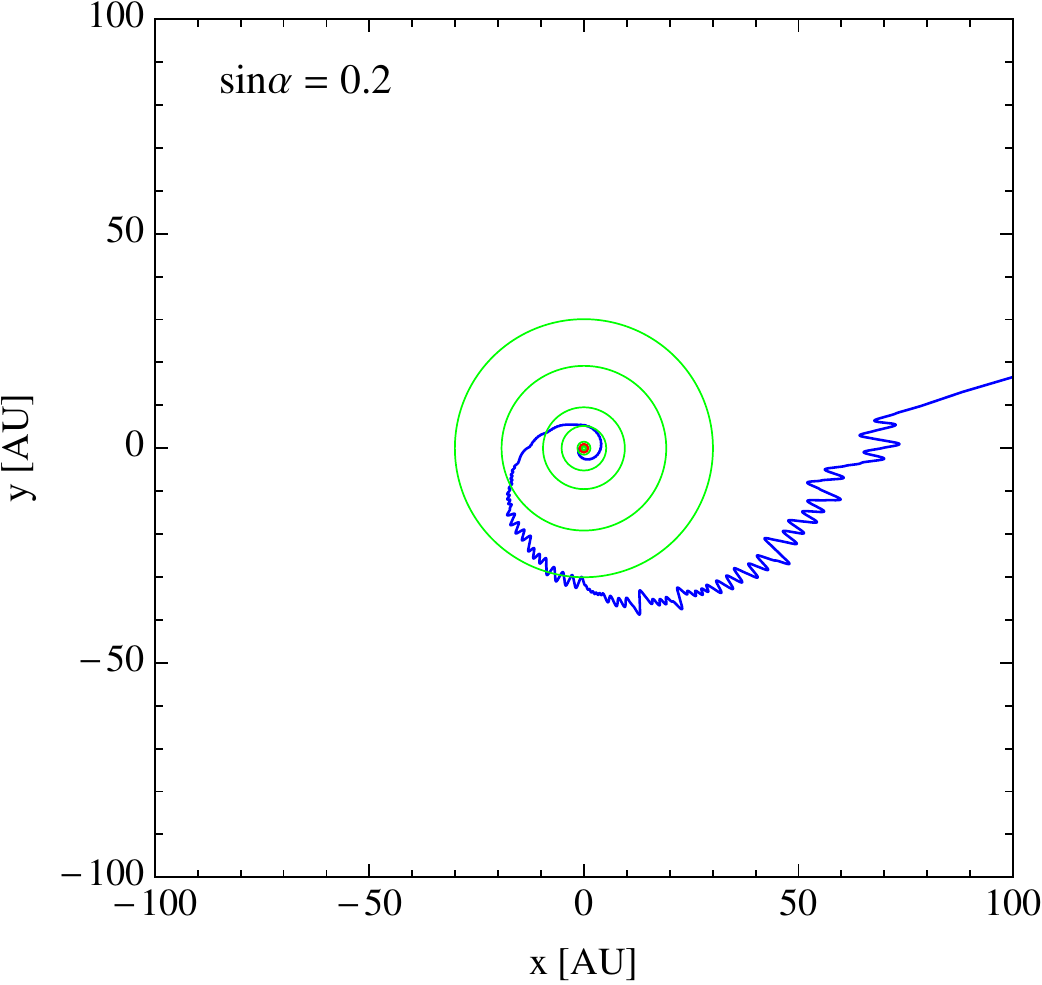}
~~~~\includegraphics[width=4.2cm]{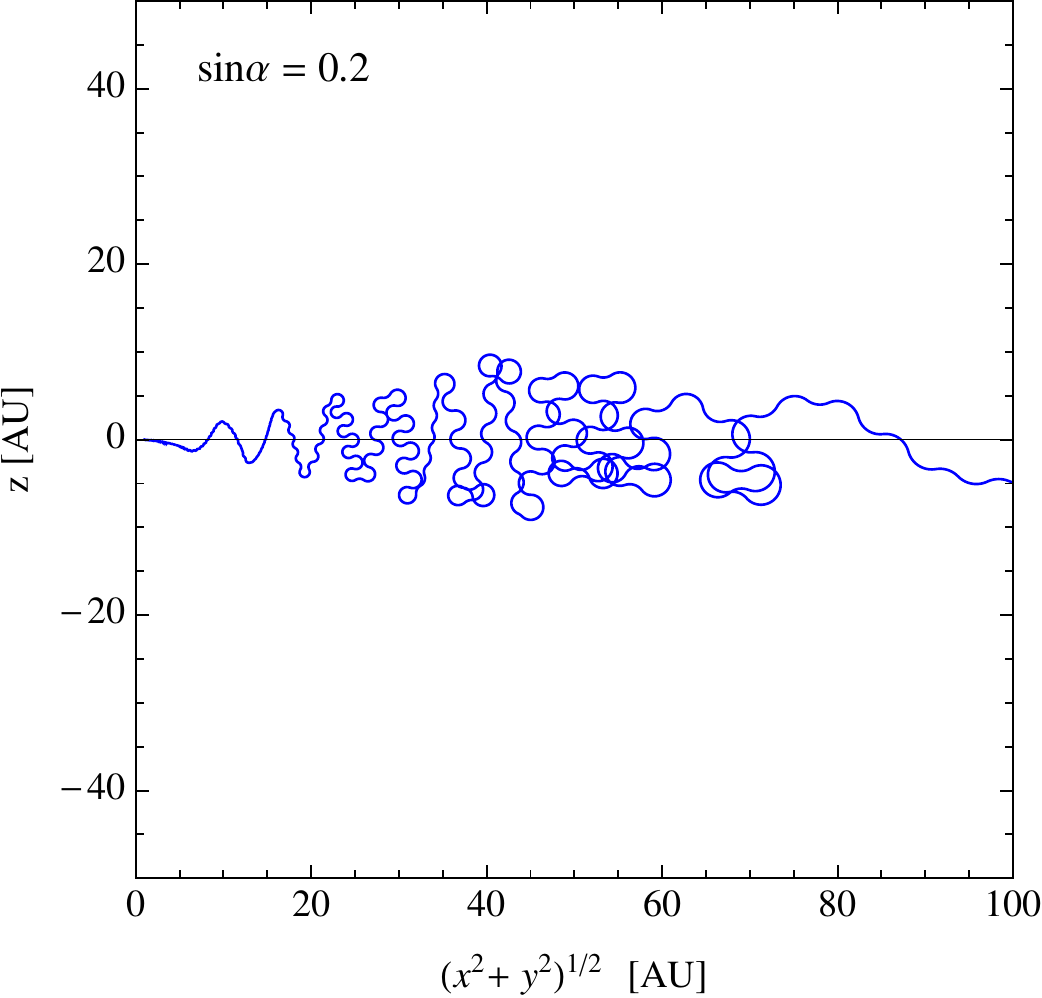}
\end{center}

\begin{center}
\includegraphics[width=4.4cm]{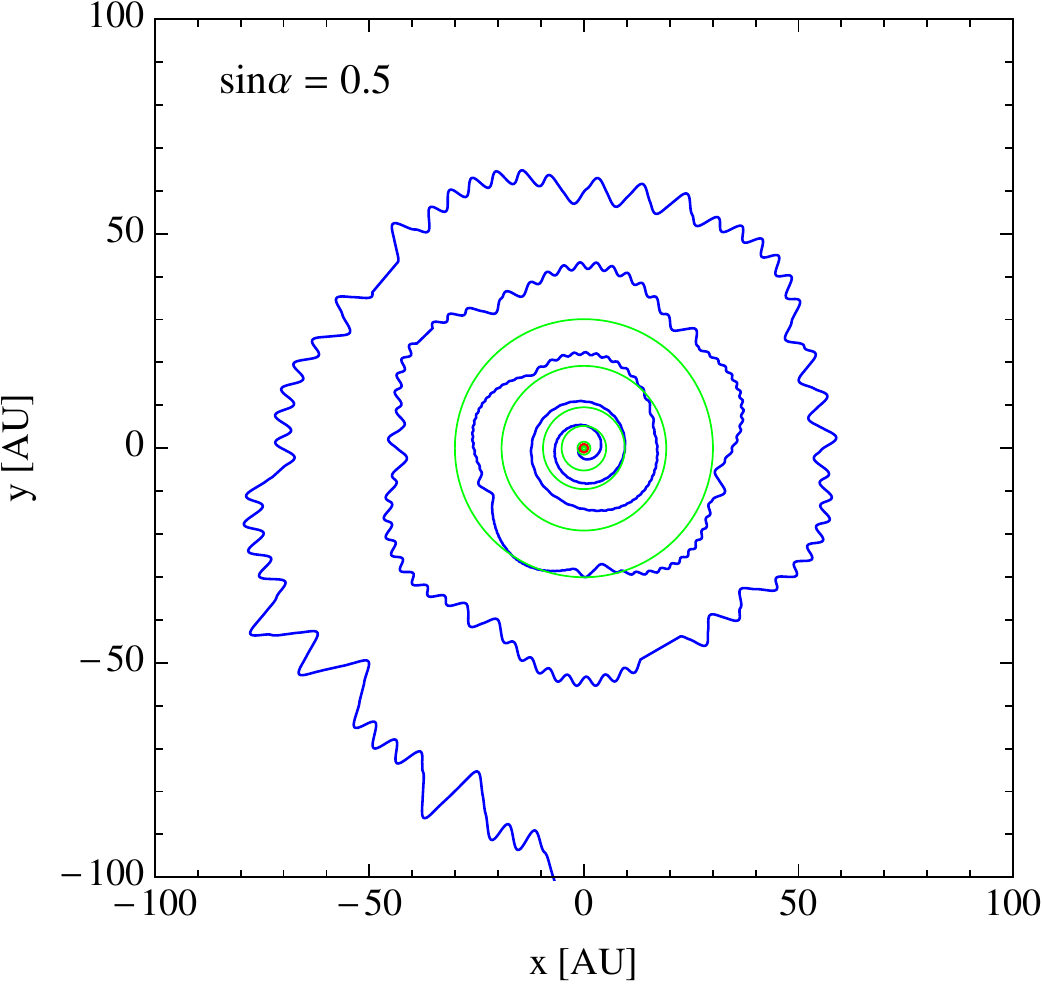}
~~~~\includegraphics[width=4.2cm]{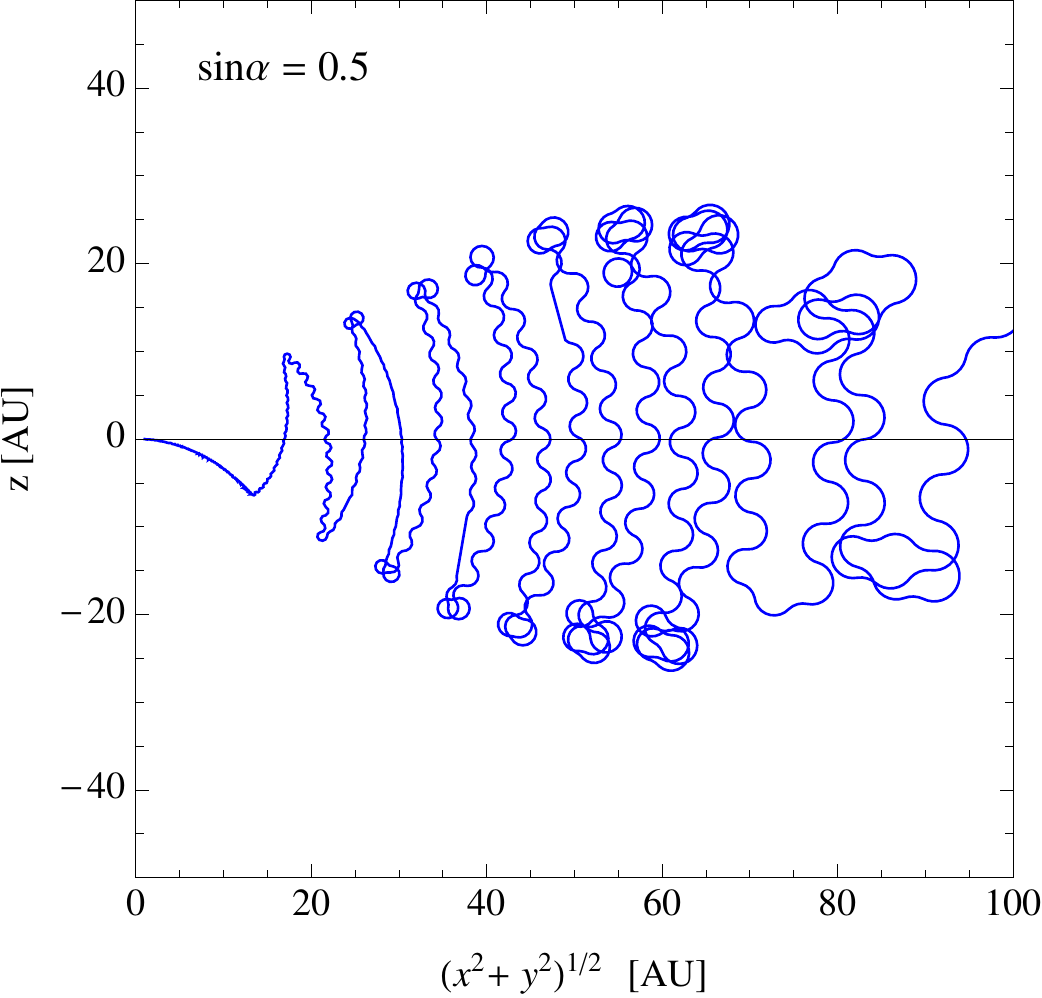}
\end{center}

\vspace{0.2cm}
\begin{center}
\includegraphics[width=4.4cm]{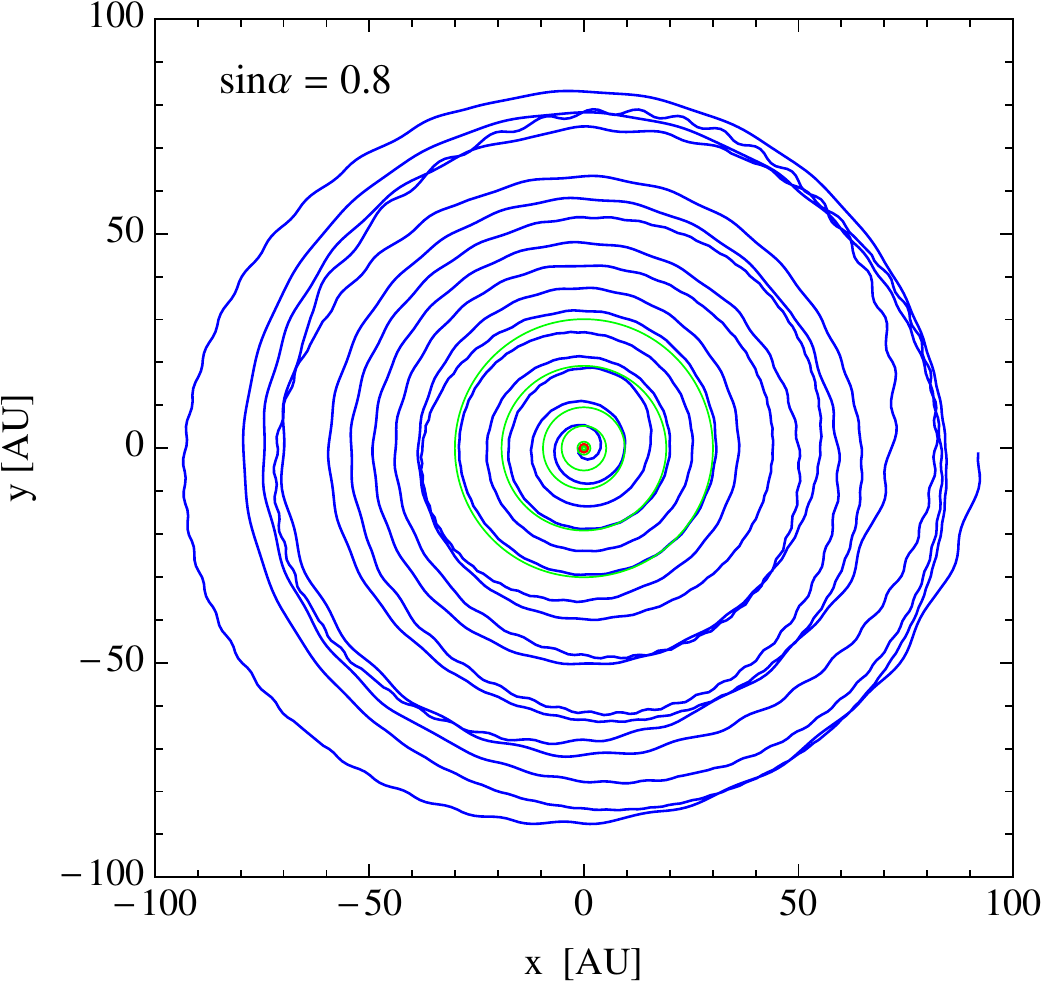}
~~~~\includegraphics[width=4.2cm]{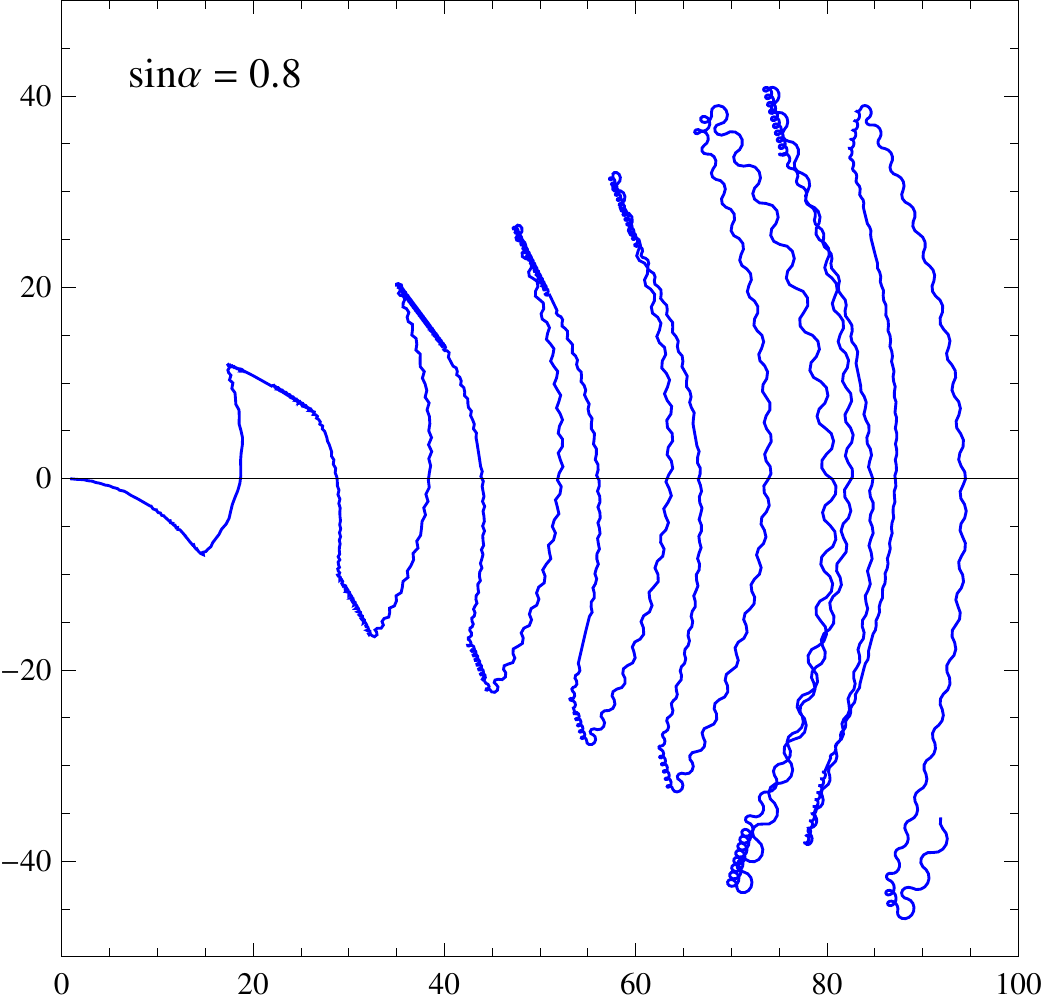}
\end{center}

\caption {\footnotesize
Projections of the past trajectory of an electron observed at the Earth 
with energy 10~GeV. The calculations are for $q A < 0$. The 
trajectory is calculated for four different values 
of the tilt angle: $\sin\alpha = 0.02$, 0.2, 0.5 and 0.8.
\label{fig:traj1}
 }
\end{figure}

The properties of the ``out'' (``in'') part of the trajectories of charged particles
for the case $q A > 0$ ($q A < 0$), depend strongly on the shape
of the HCS, and in particular on the tilt angle $\alpha$, because the
particle remains close the sheet, and follows its shape.
This is illustrated in fig.\ref{fig:traj1} that shows (for the case
$q A <0$) the past (or ``in'')
trajectories of $e^\pm$ observed at the Earth with $E = 10$~GeV.
The trajectories are shown in two projections
 ($\{x,y\}$) and ($\{\sqrt{x^2+y^2},z\}$), for four different values 
of the tilt angle ($\sin\alpha = 0.02$, 0.2, 0.5 and 0.8).
For small tilt angle (as already shown in the example of fig.~6--9) 
the past trajectory is approximately radial and confined
to the heliospheric equatorial plane.
For large tilt angle the trajectory 
undergoes oscillations around the
the equatorial plane with an amplitude that increases with $\sin\alpha$ and
$\sqrt{x^2 +y^2}$ reflecting the shape of the heliospheric current sheet;
at the same time the trajectory rotates around the $z$ axis
following the shape of the field lines.

\begin{figure} [ht]
\begin{center}
\includegraphics[width=14cm]{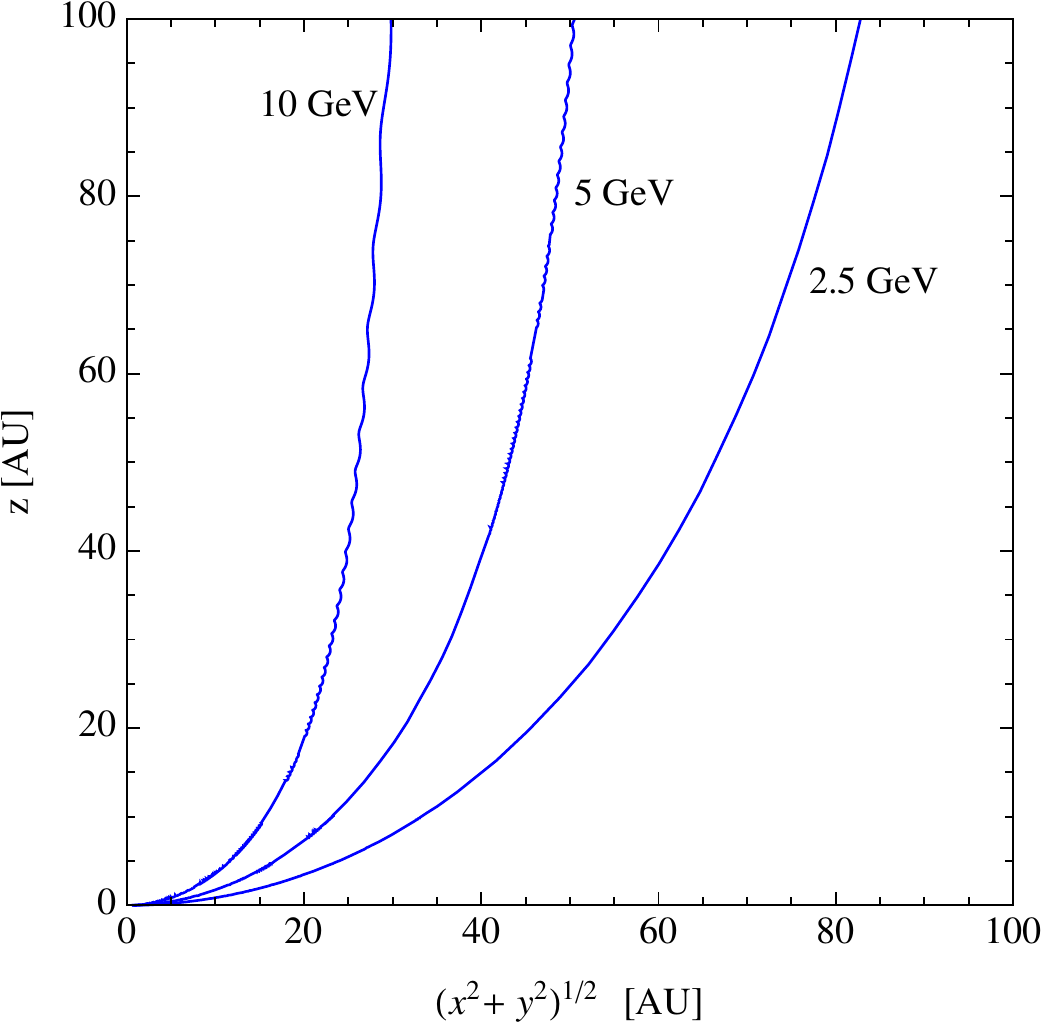}
\end{center}

\caption {\footnotesize
Projection of the ``in'' part of the space trajectories of
$e^\pm$ observed at the Earth with energy 
$E = 2.5$, 5 and 10~GeV. The calculation is for $q A > 0$.
\label{fig:projrho}
 }
\end{figure}

The spirals that form the 
``in'' trajectories when $q A > 0$ and the ``out ''trajectories when $q A < 0$
have a shape that is independent from the tilt angle of the HCS, because
they remain on one side of the HCS. 
The projection of the spiral in the $\{\sqrt{x^2 + y^2}, z\}$ plane 
has a quasi--parabolic shape that depends on the particle momentum 
as illustrated in fig.\ref{fig:projrho}, that shows
this projection for three particle momenta. The energy dependence
of the shape can be understood as the consequence of 
a drift velocity that is proportional to the particle energy.

\begin{figure} [ht]
\begin{center}
\includegraphics[width=12.0cm]{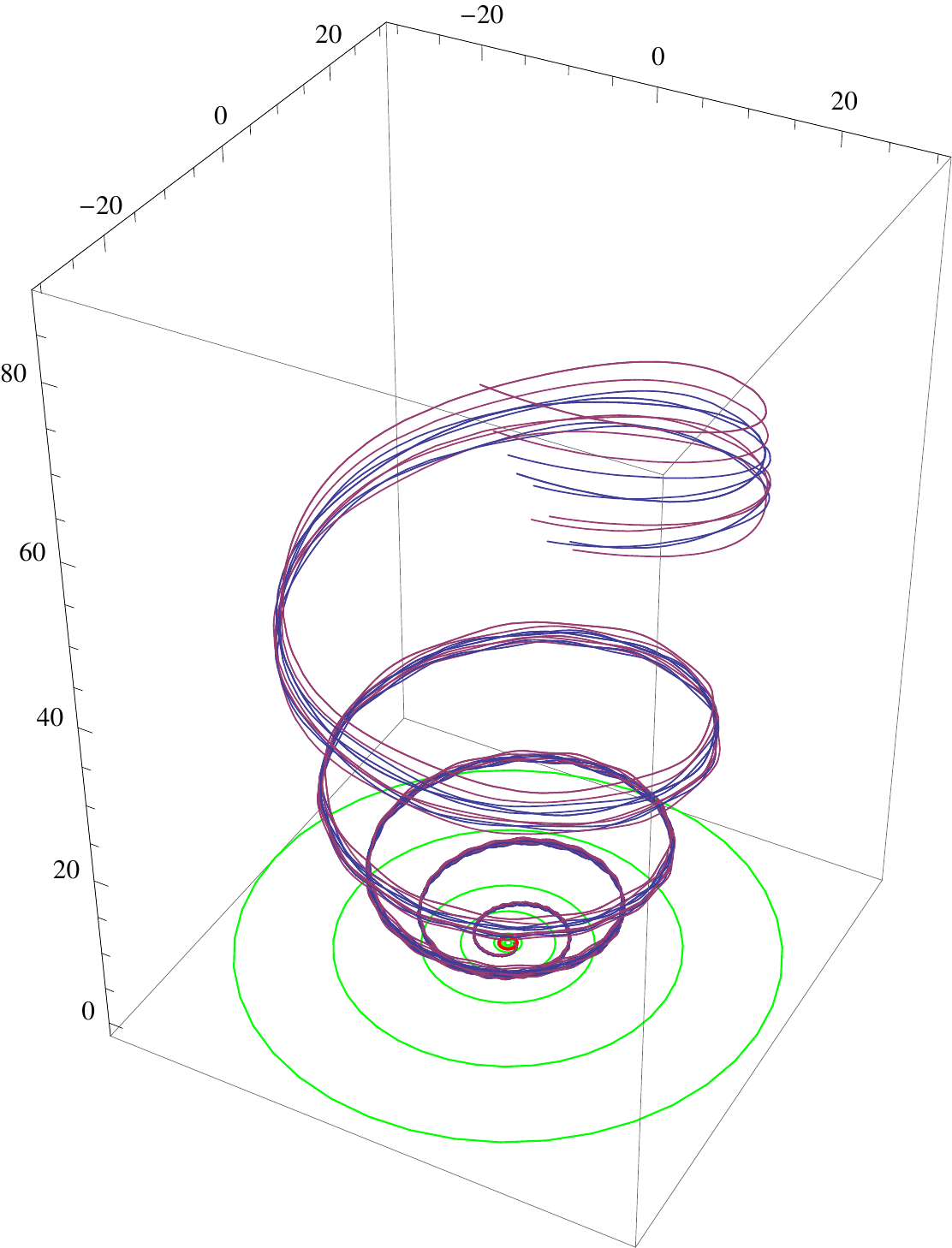}
\end{center}
\caption {\footnotesize
Past trajectories of $e^\pm$ that arrive at the Earth
with energy 12~GeV and different directions.
The calculation is valid for $q A > 0$.
\label{fig:posang}
 }
\end{figure}

\begin{figure} [ht]
\begin{center}
\includegraphics[width=12.0cm]{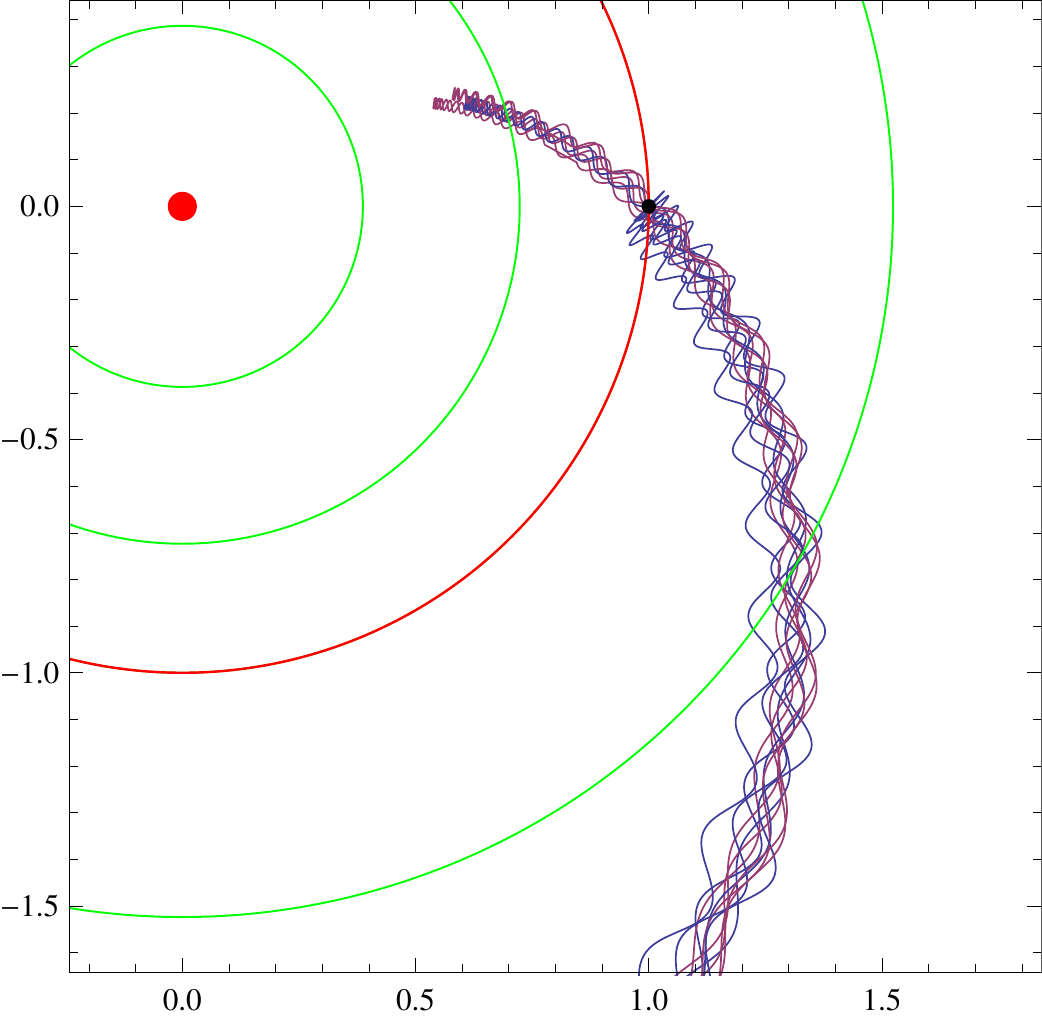}
\end{center}
\caption {\footnotesize
Projection in the equatorial plane of the
``in'' part of the trajectories of $e^\pm$ that arrive at the Earth
with energy 12~GeV and different directions.
The calculation is valid for $q A > 0$.
\label{fig:posangproj}
 }
\end{figure}

Note that the asymptotic direction (for $t \to -\infty$) of the particle 
has polar angle $\theta$ that is approximately 0 or $\pi$, 
for upgoing or down going spirals.
This property is also illustrated in fig.\ref{fig:posang} and~\ref{fig:posangproj}
that show the past trajectories
of particles observed at the Earth with 12~GeV, and different directions that span
the entire solid angle at the detector point. One can see that for $t \to -\infty$ the
directions of all particles converge to the same one 
(that is approximately $\theta = 0$).

\section{Energy Loss}
\label{sec:eloss}

\begin{figure} [ht]
\begin{center}
\includegraphics[width=12.0cm]{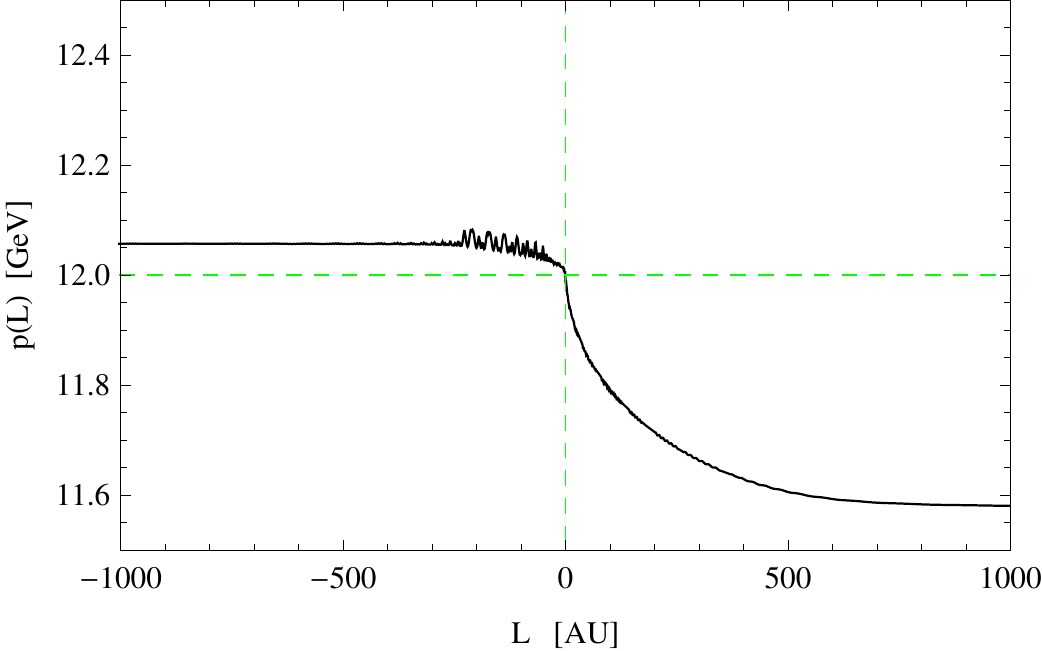}

\vspace{0.9 cm}
\includegraphics[width=12.0cm]{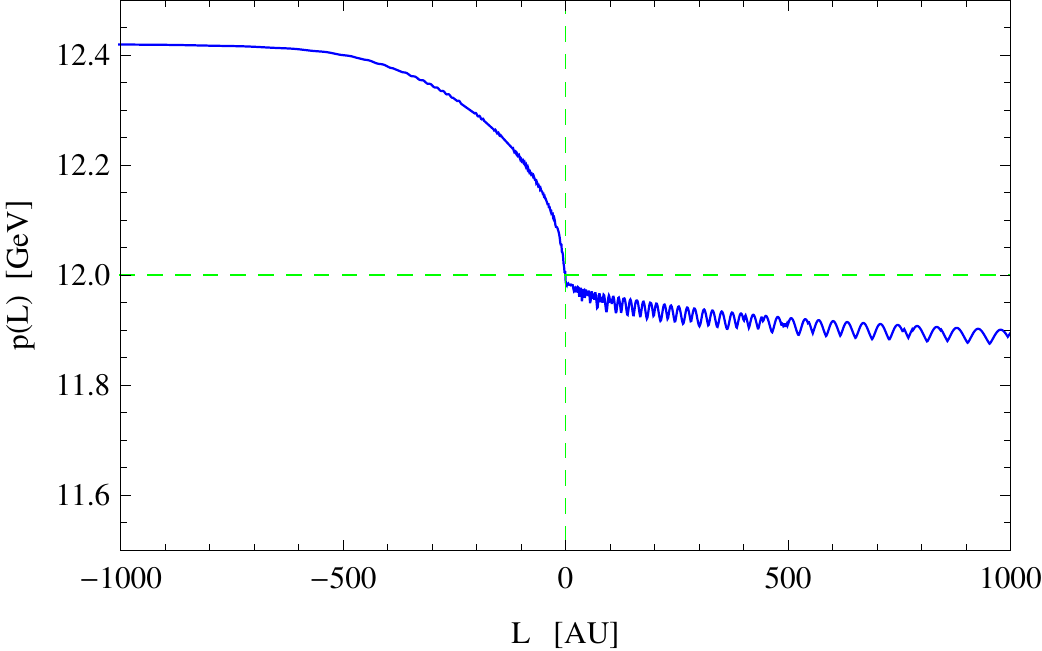}
\end{center}
\caption {\footnotesize
Energy evolution of an $e^\pm$ particle
observed at the Earth with energy 12 GeV
plotted as a function of the length 
$L = c \beta \,t$ (in astronomical units) along the trajectory.
The top (bottom) panel is for $q A > 0$ ($q A < 0$).
The space trajectory for the case $q A > 0$ 
is shown in figures~6--10.
\label{fig:sp2a}
 }
\end{figure}

\begin{figure} [ht]
\begin{center}
\includegraphics[width=12.0cm]{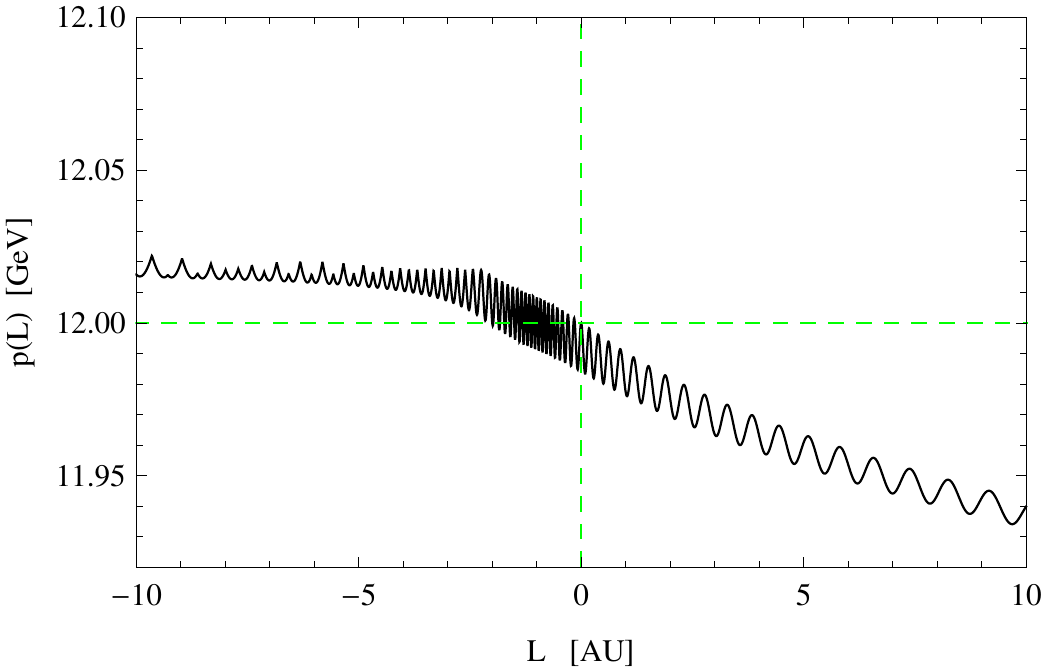}

\vspace{0.9 cm}
\includegraphics[width=12.0cm]{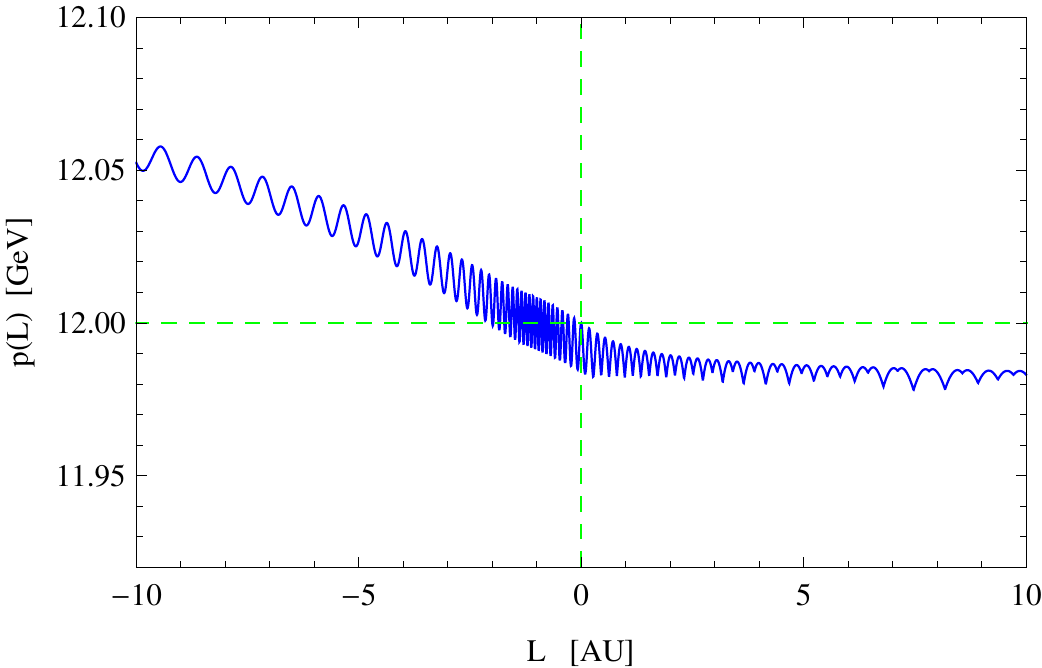}
\end{center}
\caption {\footnotesize
Energy evolution of an $e^\pm$ particle
observed at the Earth with energy 12 GeV
(the space trajectory is shown in fig.\ref{fig:example3D})
plotted as a function of the length 
$L = c \beta \,t$ (in astronomical units) along the trajectory.
The top (bottom) panel is for $q A > 0$ ($q A < 0$).
\label{fig:sp1a}
 }
\end{figure}

The equations of motion 
for the propagation of a charged particle in the heliosphere
imply an energy variation given by:
\begin{equation}
\frac{dE}{dt} = q \; \frac{d\vec{x}}{dt} \cdot \vec{\Ef} 
\end{equation}
Two examples of this energy evolution,
for an $e^\pm$ observed at the Earth with $E =12$~GeV 
are shown in fig.\ref{fig:sp2a} that plots the energy as a function
of the length along the trajectory. 
The portion of the trajectories
closest to the Earth is also shown on an enlarged in fig.\ref{fig:sp1a}.
In the example the 
parameters of the heliospheric field are:
$B_0 = 10^{-4}$~Gauss, $w = 400$~km/s and $\sin\alpha = 0.05$.

Inspecting the figures, one can see that 
after averaging for the small, fast oscillations
(produced by the gyration motion around
the magnetic field lines, that rapidly change the
orientation of the particle velocity with respect to the electric field),
the particles continuosly lose energy, both entering and exiting the 
heliosphere. 
This behavior was predicted in section~\ref{sec:drift} and with equation
(\ref{eq:elossav}). 
The initial energy of the particle $E_i = E_{\rm obs} + \Delta E$ can be
obtained as the limit of $E(t)$ for $t \to -\infty$, and 
depends on the sign of the particle.
For $q A > 0$ one has $\Delta E_+ \simeq 0.42$~GeV for $q A < 0$, 
 $\Delta E_-\simeq 0.12$~GeV.

\begin{figure} [ht]
\begin{center}
\includegraphics[width=12.0cm]{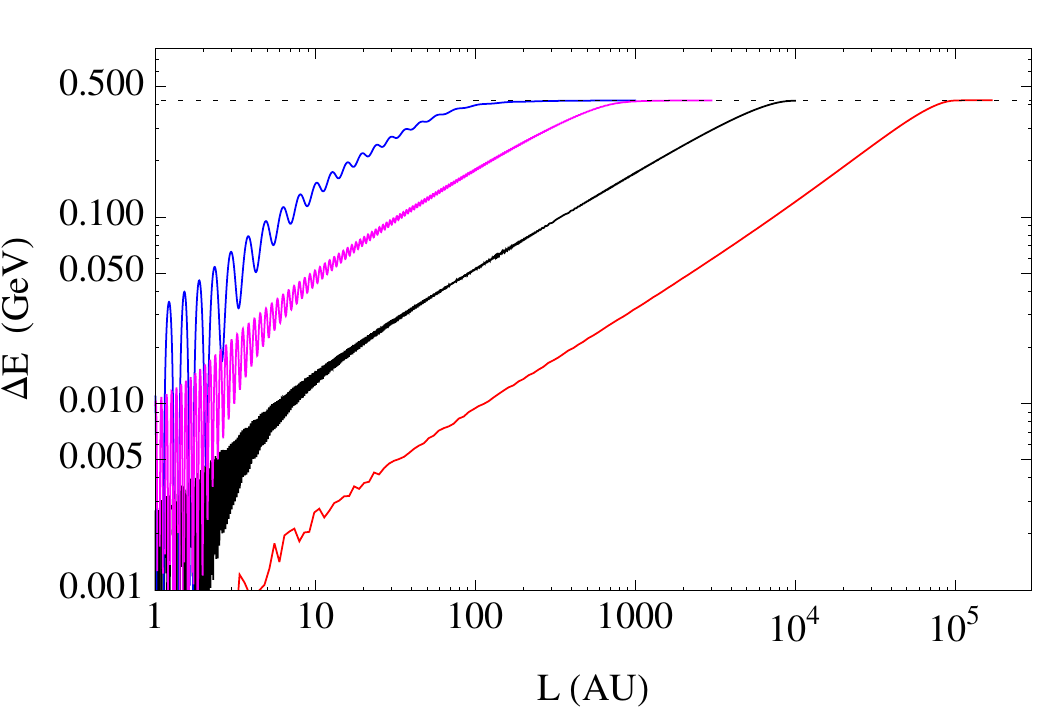}
\end{center}
\caption {\footnotesize
Time evolution (in the past) of the energy of $e^\pm$ observed at 
the Earth with energy of 1, 3, 10 and 30~GeV.
for the case $qA > 0$.
The direction of the particle at the Earth and the position of the Earth
is equal for all particles.
\label{fig:e1}
 }
\end{figure}

Figure~\ref{fig:e1} shows (for the same configuration of the 
heliospheric fields) the past time evolution of 
four particles (in the case $q A > 0$) that reach the Earth with energy
1, 3, 10 and 30 GeV. The remarkable thing, that is immediately apparent 
inspecting the figure, is that the energy loss 
suffered by all particles (in this case for $q A > 0$) 
that reach the Earth is equal, 
(and with value 0.42~GeV). 
From fig.\ref{fig:e1} one can also see that 
the time needed by a charghed particle
to reach the Earth from the boundary of the heliosphere
 has the energy dependence $E^{-1}$, reflecting a drift velocity 
proportional to $E$.
On the other hand,
as shown in equation (\ref{eq:elossav}), the energy loss $-dE/dt$ 
is proportional to $E$.
Combining the two effects one arrives to 
a total energy loss that is independent from energy.

The result that the energy loss of particles that 
reach the Earth for $q A > 0$ is a constant 
(for fixed values of $B_0$ and $w$) can be easily seen
as an elementary consequence of the existence
of an electric field potential (as given in equation (\ref{eq:vpotential})),
and of the fact that the trajectories are spirals around
the heliospheric $z$ axis. 

The energy variation of a charged particle that 
follows a trajectory $T$ can be calculated performing the line integral:
\begin{equation}
E_f = E_i + q \, \int_{T}d \vec{\ell}\cdot \vec{\Ef}(\ell) 
\end{equation}
Because of the existence of the electric potential, 
if the trajectory never crosses the heliospheric current sheet
(where the electric potential is discontinuos), the energy loss
$\Delta E = E_i - E_f$ is given by:
\begin{equation}
 \Delta E = q \left [ V(\vec{x}_f) - V(\vec{x}_i)\right ]~.
\label{eq:potential1}
\end{equation}
and is proportional to the difference between the electric
potential calculated at the initial and final point
on the trajectory.

More in general, for a trajectory that crosses $n$ times the 
heliospheric current sheet, one has:
\begin{equation}
\Delta E =  q \left [ V(\vec{x}_f) - V(\vec{x}_i) \right ] 
+q \; \left [ \sum_{k=1}^n \Delta V_{\rm disc} (\vec{x}_k ) \right ]
\label{eq:eloss1}
\end{equation}
where the summation is over all points $\{\vec{x}_k \}$ 
of intersection 
of the trajectory with the HCS, and 
$\Delta V_{\rm disc} (\vec{x}_k)$ is the potential jump 
at the discontinuity.

An alternative way to write the result (\ref{eq:eloss1}) is:
\begin{equation}
\Delta E = 
-q \; \left [ \sum_{j=1}^{n+1} \Delta V_j \right ] \;, 
\label{eq:eloss2}
\end{equation}
that is as the sum of the contributions 
as in equation  (\ref{eq:potential1})  of the ($n+1$) trajectory
segments that divide the trajectory, having the points of
discontinuity for the potential only at the boundaries.

The result (\ref{eq:potential1}) is sufficient to compute the energy
variation in the case $q A > 0$. In this case 
the charged particles reach the Earth traveling 
on a (downgoing or upgoing) spiral, starting from a very large value 
of $|z|$, and (with very few exceptions) 
arrive at the Earth without ever crossing the HCS.
The energy loss can then be estimated as:
\begin{eqnarray}
\left . \Delta E \right |_{q A>0} 
& = & q \left [  V(z=0) -
V(|z| \to \infty) \right ] \nonumber \\[0.3 cm]
&= & + q \, A \, |V_\infty| = + |q| \; |V_\infty| ~.
\label{eq:elossa}
\end{eqnarray}
In the the first step in the second line
we have used the fact that the potential 
at $z=0$ vanishes, and equation (\ref{eq:vinfty}) for the value of 
the potential at $|z|$ large.
In the final step we have used the condition $qA > 0$.
Substituting the expression of the potential at large $z$
$|V_\infty|$ given in equation (\ref{eq:vinfty}) one finds the
energy loss:
\begin{equation}
\left . \Delta E \right |_{q A>0} 
= Z \, |q_e| \; 
B_0 \, \frac{\Omega \, r_0^2}{c} 
\simeq 0.422\; Z \;
\left (\frac{B_0}{10^{-4}~{\rm Gauss}} \right )~{\rm GeV}~.
\label{eq:numeric1}
\end{equation}

It is less straightforward to estimate the energy loss 
for the case $q A <0$. In this case the
magnetic drift pushes the particle away from the heliospheric current sheet,
and therefore the particles can arrive at the Earth 
only traveling close the heliospheric equator.
Since the points  $\vec{x}_f$ (the Earth position)
and $\vec{x}_i$ (where the particle enters the heliosphere) 
have both $z$ coordinate close to zero,
one has $V(\vec{x}_i) \simeq V(\vec{x}_f) \simeq 0$, 
and the first term in the right hand side of equation (\ref{eq:eloss1}) 
vanishes. On the other hand the particles traveling close to the 
HCS will cross it several times, and the second term in the equation
(that sums the potential jumps at the discontinuities) contributes
to the energy variation.

It is easy to see that also in this case one has an 
energy {\em loss}, because the drift is always against the direction of the
electric force. The energy loss  for the case $q A < 0$
is  however sensitive to the value of the HCS tilt angle, 
and in  fact depends approximately linearly on $\sin\alpha$.

Having defined the electric potential  with the 
convention  that $V =0$ for points on the heliospheric equator, 
one has  the potential   on the two  sides of the 
HCS are equal and opposite.  Using equation (\ref{eq:vpotential})  one  has:
\begin{equation}
V_{\pm} = \mp A \, |V_\infty| \; \frac{z}{r}
\end{equation}
The points on the sheet  have  $|z|/r$  in the  interval $0 \le |z|/r \le \sin\alpha$,
so that the  each crossing of the HCS gives a contribution to the
energy loss  in the interval:
\begin{equation}
0 \le \Delta E_k =
q \; \Delta V_{\rm disc} (\vec{x}_k) \le  2 |q| \, |V_\infty| \, \sin\alpha ~.
\label{eq:deltak}
\end{equation}
The points  on the HCS that  satisfy the condition $V = \pm |V_\infty| \sin \alpha$
are those on the crests of the waves in the ondulating HCS
(see fig.\ref{fig:sheet_section}).
Equation (\ref{eq:deltak})  implies that for small tilt angle,
when the HCS nearly coincides with the equatorial plane,
all $\Delta E_k$  (and therefore $\Delta E$)  are small.

\begin{figure} [ht]
\begin{center}
\includegraphics[width=12.0cm]{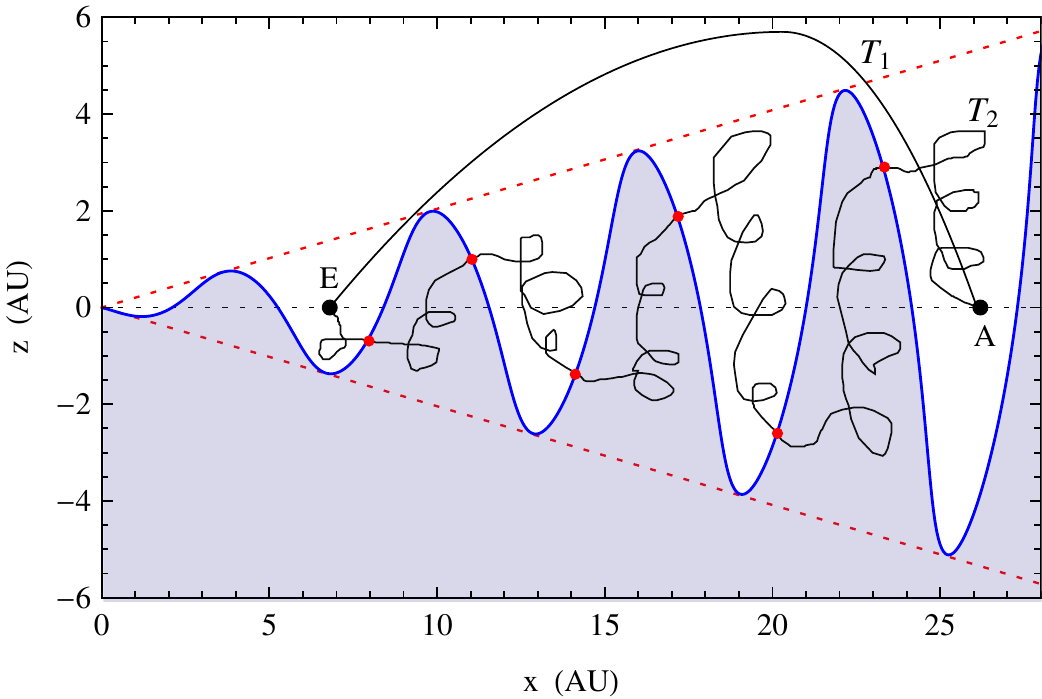}
\end{center}
\caption {\footnotesize
The figure shows an $\rperp z$ section of the heliosphere. The wavy line
is the heliospheric current sheet that separates regions of space 
where the field lines enter and exit the page. Note that the points 
on the current sheet that also satisfy the condition 
$z = \sin\alpha \, x$ (or 
$z = -\sin\alpha \, x$) have equal potential. 
Points $A$ and $E$ have the same potential. The line integral of the 
electric field along the trajectory $T_1$ is zero, while the 
line integral along $T_2$ that crosses several times the current sheet
is non vanishing.
\label{fig:section}
 }
\end{figure}

For a visualization of the energy loss in the case $q A < 0$ one can inspect
fig.\ref{fig:section} that illustrates schematically 
the propagation of a charged particle   (in the case $q A <0$)
from a  point A  on  the heliospheric equator to point E (the Earth). 
Both the starting and final points
of the trajectory are at $z=0$ and have electric potential ($V=0$),
therefore  the line integral of the electric field 
along any trajectory (like $T_1$ in the figure) that
does not cross the HCS vanishes.
On the other hand the dynamically possible trajectories that connects the 
points A and E, looks schematically  like $T_2$. 
The magnetic drift pushes the particle
toward $+z$ ($-z$) when the particle is above (below) the HCS,
however the trajectory meets and crosses several time
the HCS because of the waviness of its shape. 
In each segment of the trajectory between HCS crossings,
the particle loses energy because it moves against the electric force
that always points in a direction opposite to the drift, 
that is toward $-z$ ($+z$) above (below) the HCS,
so that the particle  continuously loses energy.

The total energy loss of the particles
that reach the Earth in the case  $q A < 0$ is not unique,
because it depends on how many times
and at what points a particle crosses the HCS. Numerical studies 
indicate however that the loss has  a surprisigly small dispersion.
As it is intuitive looking at equation
(\ref{eq:deltak}) the average energy loss is proportional to the 
product $|V_\infty| \; \sin\alpha$, so  that
\begin{equation}
\left .\left \langle \Delta E \right \rangle \right |_{q A<0} 
=  f ~|q| \; |V_\infty| \; \sin \alpha 
\label{eq:elossb}
\end{equation}
Numerical experiments suggest a value $f \simeq 1.5$--2.5.
 
The results (\ref{eq:elossa}) and (\ref{eq:elossb}) can be 
summarized as:
\begin{equation}
\Delta E \simeq 
 \begin{cases} 
|q| \, |V_\infty| 
&  \mbox{for  }  q A > 0 ~, \\[0.2 cm]
|q| \, |V_\infty| \; f \, \sin \alpha & \mbox {for } q A < 0~.
\label{eq:eloss-fin}
\end{cases}
\end{equation}
The quantity $f$ is of order 2, 
and the potential of $V_\infty$, given in equation (\ref{eq:vinfty}),
depends on the absolute normalization of the  heliospheric magnetic field.
Numerically one has:
\begin{equation}
|q| \, |V_\infty|  = Z \, |q_e| \; 
B_0 \, \frac{\Omega \, r_0^2}{c} 
\simeq 0.422\; 
\left (\frac{B_0}{10^{-4}~{\rm Gauss}} \right )~{\rm GeV}~.
\label{eq:numeric2}
\end{equation}
Both $B_0$  and the tilt angle $\alpha$  depend on  the solar
magnetic activity and vary with time. 

The energy loss depends 
on the sign of the electric charge, and the role of particles with
positive or negative charge is exchanged when the solar polarity
is reversed.  The energy loss is larger  for the case $q A > 0$ when
the tilt angle is small  ($\sin \alpha \lesssim 0.5$),
and is larger for the case $q A <0$ when the tilt angle is large
 ($\sin \alpha \gtrsim 0.5$).

\section{Energy Spectra}
\label{sec:energy-spectra}

The fundamental problem in the study of solar modulations is to relate
the energy spectra observed near the Earth at different times,
to one unmodulated spectrum that gives the distribution
at the boundary of the heliosphere. 
The unmodulated spectrum is assumed to be approximately stationary 
(on a time scale of order 10--100~years) 
and is commonly called the ``Local Interstellar Spectrum'' (or LIS).

In the model described here, and with the approximation that
charged particles lose a well defined energy $\Delta E_{\pm} (t)$
(given by equation (\ref{eq:eloss-fin})) the problem has a simple
closed form solution. Calling $\phi_{\rm obs} (E,t)$ the observable flux
at the Earth, and $\phi_0 (E)$ the LIS, one has:
\begin{equation}
\phi_{\rm obs} (E,t) = \phi_0 [ E + \Delta E(t) ]
~\left [\frac{E^2 - m^2}{[E + \Delta E(t)]^2 - m^2} \right ]~.
\label{eq:ffield0}
\end{equation}
This result is also the well known expression of the so called
``Force Field Approximation'' (FFA) for the solar modulations 
\cite{Gleeson:1968zza}.

The remarkable point is that in the context of the discussion here
the ``Force Field'' is real, and is due to the electric field that fills
the heliosphere, and the FFA expression of equation (\ref{eq:ffield0})
is an exact (for $q A > 0$) or nearly exact (for $q A < 0$) result.

Equation (\ref{eq:ffield0}) can be derived as 
a simple consequence of the Liouville theorem,
that states that if the Lagrangian of a physical system 
is time independent, the phase space density of an ensemble of 
particles remains constant along a particle trajectory.
More formally, denoting as $\rho(\vec{x},\vec{p},t) = d^6N/(d^3x \,d^3p)$ the phase
space density, one has that if $\{\vec{x}(t_1), \vec{p} (t_1) \}$ 
and $\{\vec{x}(t_2), \vec{p}(t_2)\}$ are two points on 
one solution of the equations of motion then:
\begin{equation}
\rho [\vec{x}(t_1), \vec{p}(t_1),t_1] =
\rho [\vec{x}(t_2), \vec{p}(t_2),t_2] ~.
\label{eq:liouville}
\end{equation}

The FFA  expression (\ref{eq:ffield0})   follows  from 
(\ref{eq:liouville})   and the assumptions that the interstellar spectrum 
at the boundary of the heliosphere is  stationary and isotropic.
The flux (differential in energy)  is  related  to the phase
space density  by the equation:
\begin{equation}
\phi(E, \hat{p}, \vec{x},t) 
\equiv \frac{dN}{dE \, \, dA \, dt \, d\Omega} =
v \; p^2 \; \frac{dN}{d^3p \, d^3 x} \; \frac{dp}{dE} =
 p^2 ~\rho(\vec{x}, \vec{p},t) 
\label{eq:phi-rho}
\end{equation}
(with $E = \sqrt{p^2 + m^2}$).
The particle velocity cancels in the last equality
because the Jacobian factor is $dp/dE= 1/v$.
From the condition that the phase densities of the 
particles at the Earth and outside the heliosphere are equal, 
one obtains a relation between the LIS and the observed spectrum.
Using equation (\ref{eq:phi-rho}) and the assumptions that the phase space
density at the boundary of the heliosphere is isotropic one then obtains:
\begin{equation}
\phi_{\rm obs} (E) \, p_i^2 = \phi_0 (E_i) \, p^2
\end{equation}
where $E$ and $p$ are the energy and momentum at the Earth, and 
$E_i$ and $p_i$ the same quantities outside the heliosphere.
Equation (\ref{eq:ffield0}) follows from the condition
$E_i = E + \Delta E$.

The FFA algorithm of (\ref{eq:ffield0}) is symmetric, and it can be used to
compute the observable flux at the Earth from the LIS spectrum, 
or viceversa, to reconstruct the unmodulated LIS spectrum from the observations.
For this purpose (dropping the time dependence of $\Delta E$ in the notation) 
one can recast the equation in the form:
\begin{equation}
\phi_{0} (E) = \phi_{\rm obs} [ E - \Delta E ]
~\left [\frac{E^2 - m^2}{[E -\Delta E]^2 - m^2} \right ]~.
\label{eq:ffield0a}
\end{equation}
In this case there is obviously a limitation,
since equation (\ref{eq:ffield0a}), even in the case of a perfect meaurement,
allows to determine the interstellar spectrum 
only above a minimum kinetic energy $\Delta E$, because for lower energy
the argument of $\phi_{\rm obs}$ in (\ref{eq:ffield0a}) becomes unphysical, and the
equation meaningless.
This simply reflects the fact that particles with a kinetic energy
lower than $\Delta E$ in interstellar space cannot reach the Earth. 

It is interesting to note that equation (\ref{eq:ffield0}) predicts 
that  in the limit of small kinetic energy 
the observable flux at the Earth $\phi_{\rm obs} (E)$ is linear in $E_{\rm kin}$.
In  fact,   in the limit  $E_{\rm kin} \to 0$
(and for $E_{\rm kin} \ll m$)  the observable flux   has 
the behavior:
\begin{equation}
\phi_{\rm obs} (m+ E_{\rm kin}) 
\simeq \phi_0 (m + \Delta E) \; \frac{2 m \, E_{\rm kin}} {\Delta E^2 + 2 m \, \Delta E} 
+ O (E_{\rm  kin}^2) 
\simeq {\rm const} \times  E_{\rm kin}~.
\label{eq:ekinsmall}
\end{equation}
As it will shown in the following,  the  linear  dependence  of the fluxes
is the observed   behavior of the fluxes,
and insures that the estimate of  the interstellar flux  
$\phi_0 (m + \Delta E)$ from equation (\ref{eq:ffield0a}) is finite.

More in general, it is possible to use expression
(\ref{eq:ffield0}) to relate spectra measured at different times.
If for example the spectrum $\phi_1(E)$, measured at time $t_1$, is 
related to the LIS spectrum by the FFA expression
with parameter $\Delta E_1$, and similarly 
the spectrum $\phi_2(E)$, measured at time $t_2$, is 
related to the LIS spectrum using the same relation 
with the parameter $\Delta E_2$, then the spectra 
 $\phi_2(E)$ can be obtained from the flux $\phi_1(E)$
with the one parameter transformation:
\begin{equation}
\phi_{2} (E) = \phi_1 [ E + \Delta E_{21}]
~\left [\frac{E^2 - m^2}{[E + \Delta E_{21}]^2 - m^2} \right ]~.
\label{eq:ffield1}
\end{equation}
where 
\begin{equation}
\Delta E_{21} = \Delta E_{2} - \Delta E_1~,
\end{equation}
or symmetrically $\phi_1(E)$ can be obtained from $\phi_2(E)$ by the same 
transformation with parameter $\Delta E_{12} = -\Delta E_{21}$.
Equation (\ref{eq:ffield1}) can be used to test experimentally the
validity of the FFA, measuring two spectra for different conditions in the heliosphere,
and studying if the spectra satisfy equation 
(\ref{eq:ffield1}) for some value of the parameter $\Delta E_{21}$.

\begin{figure} [ht]
\begin{center}
\includegraphics[width=12.0cm]{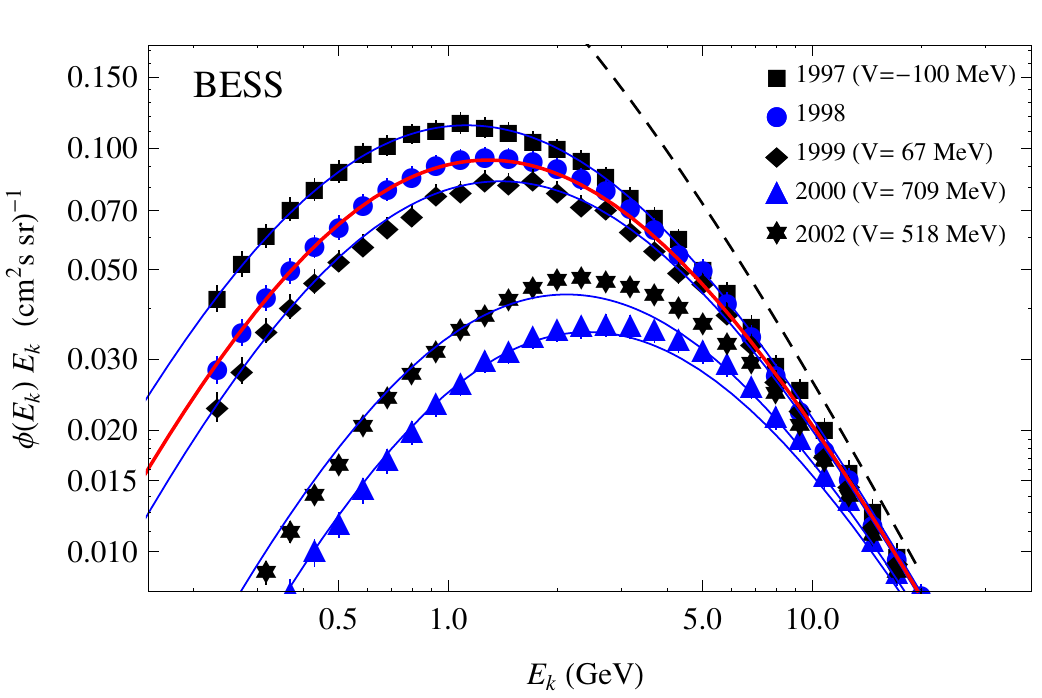}
\end{center}
\caption {\footnotesize
Proton spectra measured by the BESS collaboration at different times. 
\protect\cite{Sanuki:2000wh,Shikaze:2006je}.
The thick red line is the fit to the 1998 BESS proton given in \protect\cite{Shikaze:2006je}.
The other lines are one parameter fits to data, distorting the
curve that describes the 1998 data 
the algorithm of the Force Field approximation.
The best fit parameter of the curves is given in the figure.
\label{fig:bess}
 }
\end{figure}

\begin{figure} [ht]
\begin{center}
\includegraphics[width=12.0cm]{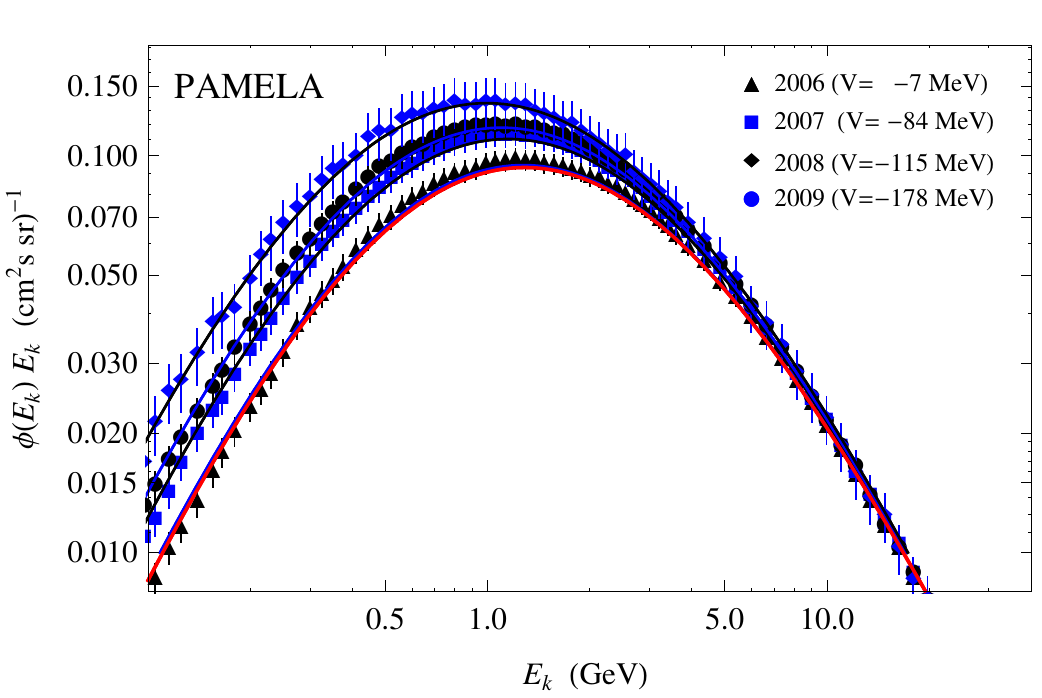}
\end{center}
\caption {\footnotesize
Proton spectra measured by the PAMELA collaboration
\protect\cite{Adriani:2013as}.
The thick red line is the fit to the 1998 BESS proton data shown in fig.\ref{fig:bess}.
The other lines are one parameter fit to the PAMELA data, where the fitting curve is the 
fit to the BESS-1998 data, distorted with the algorithm of the Force Field approximation.
The parameter of the fit is shown in the figure.
\label{fig:pamela}
 }
\end{figure}

An illustration of the FFA algorithm applied 
to measurements of the cosmic ray proton spectrum is given in fig.\ref{fig:bess} 
and~\ref{fig:pamela} that show measurements of the proton flux
measured at different times by the BESS \cite{Sanuki:2000wh,Shikaze:2006je}
and PAMELA \cite{Adriani:2013as} detectors. 
The BESS measurements were performed during (approximately one day long)
balloon flights in 1997, 1998, 1999, 2000 and 2002. 
During all this time the polarity of the heliospheric 
magnetic field was positive ($A>0$).
The measurements of PAMELA are averages of data collected in 2006, 2007, 2008 and 2009,
when the polarity of the heliospheric field was negative ($A<0$).

Fig.\ref{fig:bess} shows the five measurements of the BESS spectrometer
\cite{Shikaze:2006je}. In the figure the thick (red) line was 
chosen to give a good description of the data collected by BESS in 1998,
the other lines are calculated using the FFA algorithm of equation (\ref{eq:ffield1}) 
using the 1998 flux as starting point and fitting the
parameter $\Delta E$ for the other BESS measurements of the proton flux
performed in 1997, 1999, 2000 and 2002. 
The best fit values of $\Delta E$ (relative to the 1998 measurement) 
for the four spectra
are $-100$, 67, 709 and 518~MeV. The parameter  $\Delta E$ is positive
(negative) when the fitted spectrum is
lower (higher) than the reference flux.

Figure~\ref{fig:pamela} shows the measurements of the proton flux obtained by 
PAMELA \cite{Adriani:2013as} in 2006, 2007, 2008 and 2009.
The lines in the figure are one parameter fits to the four spectra
with the FFA algorithm of equation (\ref{eq:ffield1}),
calculated using as starting flux the parametrization of the
measurement performed by BESS in 1998.
The best fit values for $\Delta E$ are $-7$, $-84$, $-115$ and $-178$~MeV
(in all years the flux measured by PAMELA is higher than the one measured
by BESS in 1998). The FFA fits give a very good description of the
proton spectra.

In summary, the measurements of the proton flux performed by
BESS and PAMELA during nine years (that include an inversion of the solar
magnetic phase) exhibit time variations that are reasonably well described
by the Force Field Approximation algorithm. 

In the discussion above, we have studied the validity of the 
FFA algorithm comparing measurements performed at different times,
without attempting to reconstruct the unmodulated interstellar flux.
Previous studies  have used use the  approach to  relate the observed flux
to the interstellar spectrum.
For example, in \cite{Shikaze:2006je}
the BESS collaboration  have shown 
the time dependence of their proton flux measurements also using the 
FFA algorithm, but starting from a model of the unmodulated LIS spectrum.
This LIS spectrum
(shown in fig.\ref{fig:bess} as a dashed line), 
is then used to estimate an energy loss $\Delta E$ 
(relative to the unmodulated flux) for each proton flux measurement.
The two procedures (the one performed by the BESS collaboration,
and the one discussed in this work) are perfectly consistent, because one has:
\begin{equation}
\Delta E_{j,1998} =
\Delta E_{j,{\rm LIS}} -
\Delta E_{1998,{\rm LIS}}~,
\end{equation}
where $j$ is an index that runs through the five measurements,
$\Delta E_{j,1998}$ is the energy loss with respect to the 1998 measurement
estimated in this  work,
and $\Delta E_{j,{\rm LIS}}$ the energy loss with respect to the LIS spectrum
estimated by the BESS collaboration in \cite{Shikaze:2006je}.
In the BESS study 
 the total energy loss of protons at the time of
the 1998 measurement was estimated as 591~MeV.

The point we want to make here is that it is possible to verify
the validity (or non--validity) of the FFA without making any assumption
on the shape and properties of the interstellar flux,
simply comparing measurements taken at different times, 
and testing if they are consistent
with the one parameter transformation of equation (\ref{eq:ffield1}).
If this test is succesful, then one can conclude that all measurements
are the results of the time dependent modulations of a unique
interstellar flux described by the FFA algorithm.

Having established the validity (or approximate validity) of the
FFA, the reconstruction of the unmodulated flux requires
additional theoretical assumptions, because the 
FFA algorithm has in fact infinite solutions,
parametrized by $\Delta E_{\rm LIS}$,  that are mathematically
all equally valid. To select the physically correct solution there
are in principle two different methods. 
In the first, one starts from a theoretical prediction 
for the shape of the interstellar energy spectrum,
and then use equation (\ref{eq:ffield0a})
to determine  $\Delta E_{\rm LIS}$, as ther parameter
that reproduces the observed spectrum, starting from the
interstellarone.
This method allows to  obtain an estimate of 
the total size of the modulation effects from the cosmic ray flux measurement.

The  alternative approach is to have a complete model for propagation
in the heliosphere that allows to compute from first principles
the value of $\Delta E_{\rm LIS}$ at different epochs.
Inserting this value into equation (\ref{eq:ffield0a}) one 
can then calculate the of the interstellar energy spectrum 
from the observations of the spectral shape at the Earth.

The main point want to make in this section
is that the FFA algorithm is quite successful in describing
the time variations of the cosmic rays fluxes.
This implies that the cosmic rays that arrive at the Earth
lose a (time dependent) $\Delta E$ that is approximately 
equal for all energies and has a small dispersion. 

In general  one  can expect that the relation
between the observed and the interstellar flux
is more complicated that the FFA algorithm of equation (\ref{eq:ffield0}).
For example, the energy variation $\Delta E$ could be 
a function of $E$ (or equivalently of $E_i$). 
Also in general the  energy loss of particles that traverse the heliosphere
will not be unique, but can have a distribution with a certain (possibly
energy dependent) width and shape.
The point os that the approximate validity of the FFA is 
non trvila result, that gives important constraints 
on the construction of a model of the solar modulations.

\section{Summary and outlook}
In this work we have studied the propagation of charged particles in the heliosphere
assuming only the presence of the regular magnetic field, completely neglecting 
the random magnetic field generated by turbulent motions
in the heliospheric plasma.
The presence of a magnetic field in the   solar wind,
a moving plasma with good electrical conductivity,
generates an electric field that plays an important role in 
the propagation of charged particles  in the heliosphere, and 
results in significant solar modulations.

The combined effect of the large scale heliospheric magnetic and electric fields
is that charged particles that reach the Earth from   the boundary
of the heliosphere always lose energy.
The energy loss  $\Delta E$,
to a good approximation, is independent from the energy and direction
of the particle, and is determined  only by the structure of the heliospheric fields.
The energy loss is proportional to the absolute value of the 
electric charge of the particle, but depends on the sign of the charge. 

It might appear surprising that the same field configuration can
generate energy losses for particles of opposite electric charge, but 
this is possible because particles of opposite charge reach the Earth traveling on
different trajectories.
In all cases the electric field results in a force that, 
after averaging over fast gyrations around the magnetic field lines,
is anti--parallel to the particle velocity, and causes an energy loss. 

The shape and properties of the trajectories
are determined by the structure of the heliospheric magnetic field, with the 
electric field that acts effectively only as a small perturbation.
When the product $q A$ of the particle electric charge and the solar magnetic polarity
is positive  particles  arrive at the Earth  traveling along spirals
around the heliosphere polar axis, in the opposite case  the particles  arrive
traveling close to the heliospheric current sheet (HCS) and the the heliospheric equator.

The energy loss is always linear with the value pof the magnetic field,
but in the case  $q A <0$ it also depends  on the tilt angle of the HCS,
and is  also linear with $\sin \alpha$.  
The energy loss is larger for $q A > 0$ when the tilt angle
of the HCS is small ($\sin \alpha \lesssim 0.5$), 
and it is larger for $q A < 0$ when the tilt angle is large
($\sin \alpha \gtrsim 0.5$).
A quantitative estimate of the energy loss is   given in 
equations (\ref{eq:eloss-fin}) and (\ref{eq:numeric2}), and is of order
of a few hundred  MeV.

All the results discussed in this work have been obtained for a very simple model 
of the heliosphere (a magnetic field that is a Parker spiral, 
and a solar wind that is purely radial,  and of constant velocity), 
but one can expect that they will remain valid for a more general description
of the heliospheric magnetic field and of the solar wind.


In conclusion, in this work we have demonstrated that
the deterministic  propagation of charged particles  in the heliosphere
results in an energy loss  that is sufficient to
generate important solar modulations.

Modern calculations of solar modulations effects do  include a description
of the regular heliospheric magnetic field, 
but only as a  convection term associated to the magnetic drift velocity. 
The drift velocity inverts its  direction for the transformation $q \to -q$, 
and introduce a dependence on the sign of the electric charge.
This treatment does not take into account properly
of the important fact, outlined in this work, 
that the heliosphere contains a regular electric field,
with a large scale structure.
In the standard  treatment of the solar modulations 
particles lose energy adiabatically, that is they suffer an energy 
loss proportional to energy and to the divergence of the
wind velocity ($-dE/dt \propto E \, \nabla \cdot \vec{w}$). This implies
that the total energy loss of a particle
observed at the Earth is proportional to the time that a particle
takes to penetrate the heliosphere, independently from the characteristics
of the trajectories.

The  phenomenological success of the Force Field Approximation
(as discussed in section~\ref{sec:energy-spectra})
suggests that charged particles  penetrate the heliosphere
losing an amount of energy  $\Delta E$ that is approximately  independent 
from $E$ and has a small dispersion.
This is a suprising result
when seen in the framework of the Parker equation, because it implies
that the time needed by a particle to reach the Earth has the energy dependence
$t \propto E^{-1}$, and therefore that the diffusion coefficient (that 
describes the effects of the turbulent magnetic field) 
has the energy dependence $D(E) \propto E$.
This  energy dependence  of the diffusion coefficient 
is not seen in other  astrophysical environments,
for example when studying the escape of cosmic rays from the Milky Way, where
one estimate a power law dependence $D(E) \propto E^\delta$ with $\delta = 0.3$--0.6.
Also the apparent narrowness of the $\Delta E$ distribution appears to be 
problematic.

These difficulties could perhaps be solved assuming that
a large fraction of the energy loss for cosmic rays 
that travel in the heliosphere is caused by 
a large scale electric field that permeates the solar wind.
The existence of this electric field emerges as a simple and well known
magnetohydrodynamical effect when a conducting and moving fluid 
(such as the solar wind) contains a magnetic field.
The investigation of this hypothesis clearly requires 
detailed numerical studies of the propagation of particles
in the heliosphere that take into account the presence of both
the regular and random components of the electromagnetic fields.

The existence of the regular electromagnetic fields in the heliosphere
has  very important consequences for studies on anisotropy 
of cosmic rays at the Earth, because it  clearly has effects on the 
angular distribution of particles observed at the Earth.
 his problem will be discussed in a separate work \cite{inpreparation}.

In conclusion, the results outlined in this work point to the need of 
a profound revision of the treatment of solar modulations that properly include
the regular electromagnetic fields present in the heliosphere.

\clearpage

\clearpage


\end{document}